\documentclass[aps,prc,twocolumn,showpacs,amsmath,amssymb]{revtex4-1}
\usepackage[T1]{fontenc}
\usepackage{graphicx}
\usepackage{dcolumn}
\usepackage{bm}

\begin{document}

\title{Modeling Hybrid Stars with an SU(3) Sigma Model}

\author{Rodrigo Negreiros}
 \email{negreiros@fias.uni-frankfurt.de}
\affiliation{FIAS, Goethe University, Ruth Moufang Str. 1
        60438 Frankfurt am Main, Germany}

\author{V.A. Dexheimer}
 \email{vantoche@gettysburg.edu}
\affiliation{Gettysburg College, Gettysburg, USA}

\author{S. Schramm}
 \email{schramm@th.physik.uni-frankfurt.de}
\affiliation{CSC, FIAS, ITP, Johann Wolfgang Goethe University, Frankfurt am
Main, Germany}

\date{\today}

\begin{abstract}
We study the behavior of hybrid stars  using an extended
hadronic and quark SU(3) non-linear sigma model. The degrees of
freedom change naturally, in this model, from hadrons to quarks
as the density/temperature increases. At zero temperature, we reproduce
massive neutron stars containing a core of hybrid matter of 2
km for the non-rotating case and 1.18 km  and 0.87 km, in the
equatorial and polar directions respectively, for stars
rotating at the Kepler frequency (physical cases lie in
between). The cooling of such stars is also analyzed.
\end{abstract}


\maketitle

\section{Introduction}

As the densest bodies in the universe, neutron stars are one of the best suited
candidates to look for quark matter. Studying
when and how the deconfinement from hadronic to quark matter
occurs is crucial for the understanding and identification of
such phenomena. Usual approaches for hybrid neutron stars
consist of two different models with separate equations of
state for hadronic and quark phases (see e.g.
\cite{Heiselberg1993}), connected at the point for which the
pressure of the quark phase exceeds that of the hadronic one.
Within our approach we employ a single model for the hadronic
and quark phases, avoiding the need for two separate equations of
state. This approach, that presents a more natural transition from hadronic to
quark matter, allows us to follow the spin evolution of a hybrid star with a
single equation of state. In this paper we will investigate the structural
changes that follows the stellar spin-down and the subsequent effects on the
cooling of the object.

The SU(3) non-linear sigma model introduces  baryons and quarks as flavor-SU(3)
multiplets. Baryons and quarks obtain their masses through their coupling
to the scalar fields of the theory (with an additional
coupling to the Polyakov loop as discussed below) via spontaneous  symmetry
breaking. We include the quark degrees of freedom in the hadronic model in
analogy to the PNJL model \cite{Fukushima2004} by
using an effective field that can be related to the QCD Polyakov loop,
defined via
$\Phi=\frac13$Tr$[\exp{(i\int
d\tau A_4)}]$, where $A_4=iA_0$ is the temporal component of
the SU(3) gauge field. The effect of the
field is to suppress quarks in the low density/temperature regime
and baryons at high densities and temperatures, respectively. 

This paper is
divided as follows: in section II we review the properties of the model used
for the composition and equation of state; in section III we present our
results which encompass the structure of rotating and spherically symmetric
compact stars, and cooling effects; and in section IV our conclusions are
presented.

\section{The Model}

The Lagrangian density for the sigma-type model, in mean
field approximation, is given by
\begin{eqnarray}
&L = L_{Kin}+L_{Int}+L_{Self}+L_{SB}-U,&
\end{eqnarray}
where besides the kinetic energy term for hadrons, quarks, and leptons,
the terms
\begin{eqnarray}
&L_{Int}=-\sum_i \bar{\psi_i}[\gamma_0(g_{i\omega}\omega+g_{i\phi}\phi+g_{i\rho}\tau_3\rho)+M_i^*]\psi_i,\nonumber&\\&
\end{eqnarray}
\begin{eqnarray}
&L_{Self}=-\frac{1}{2}(m_\omega^2\omega^2+m_\rho^2\rho^2+m_\phi^2\phi^2)\nonumber&\\&
+g_4\left(\omega^4+\frac{\phi^4}{4}+3\omega^2\phi^2+\frac{4\omega^3\phi}{\sqrt{2}}+\frac{2\omega\phi^3}{\sqrt{2}}\right)\nonumber&\\&+k_0(\sigma^2+\zeta^2+\delta^2)+k_1(\sigma^2+\zeta^2+\delta^2)^2&\nonumber\\&+k_2\left(\frac{\sigma^4}{2}+\frac{\delta^4}{2}
+3\sigma^2\delta^2+\zeta^4\right)
+k_3(\sigma^2-\delta^2)\zeta&\nonumber\\&+k_4\ \ \ln{\frac{(\sigma^2-\delta^2)\zeta}{\sigma_0^2\zeta_0}},&
\end{eqnarray}
\begin{eqnarray}
&L_{SB}= m_\pi^2 f_\pi\sigma+\left(\sqrt{2}m_k^ 2f_k-\frac{1}{\sqrt{2}}m_\pi^ 2 f_\pi\right)\zeta,\nonumber&\\&
\end{eqnarray}
represent the baryon (and quark) -- meson interactions, meson self-interactions,
and an explicit chiral symmetry breaking term that is responsible for producing
the 
masses of
the pseudo-scalar mesons. We will discuss the potential $U$  further along
this paper. The model has a SU(3) flavor symmetry, and the index $i$
denotes the baryon octet and the three light quarks. In our calculations we
take into account the following mesons: the vector-isoscalars $\omega$ and
$\phi$, the vector-isovector $\rho$,
the scalar-isoscalars $\sigma$ and $\zeta$ (non-strange and strange
quark-antiquark states, respectively), and
the scalar-isovector $\delta$.
The coupling constants of the model can be found in reference
\cite{Dexheimer2010}.
They were fitted to reproduce the vacuum
masses of the baryons and mesons, nuclear saturation properties (density
$\rho_0=0.15$ fm$^{-3}$, binding energy per nucleon $B/A=-16.00$ MeV, nucleon
effective mass $M^*_N=0.67$ $M_N$, compressibility $K=297.32$ MeV), asymmetry
energy ($E_{sym}=32.50$ MeV), and reasonable values for the hyperon potentials
($U_\Lambda=-28.00$ MeV, $U_\Sigma=5.35$ MeV, $U_\Xi=-18.36$ MeV). The vacuum
expectation values of the scalar mesons are constrained by
reproducing the pion and kaon decay constants. A detailed discussion of the
purely hadronic part of the Lagrangian can be found in
\cite{Papazoglou1998,Papazoglou1999,Dexheimer2008a}.

\begin{figure}
\centering
\vspace{1.0cm}
\includegraphics[width=8.5cm, height =  6.0 cm]{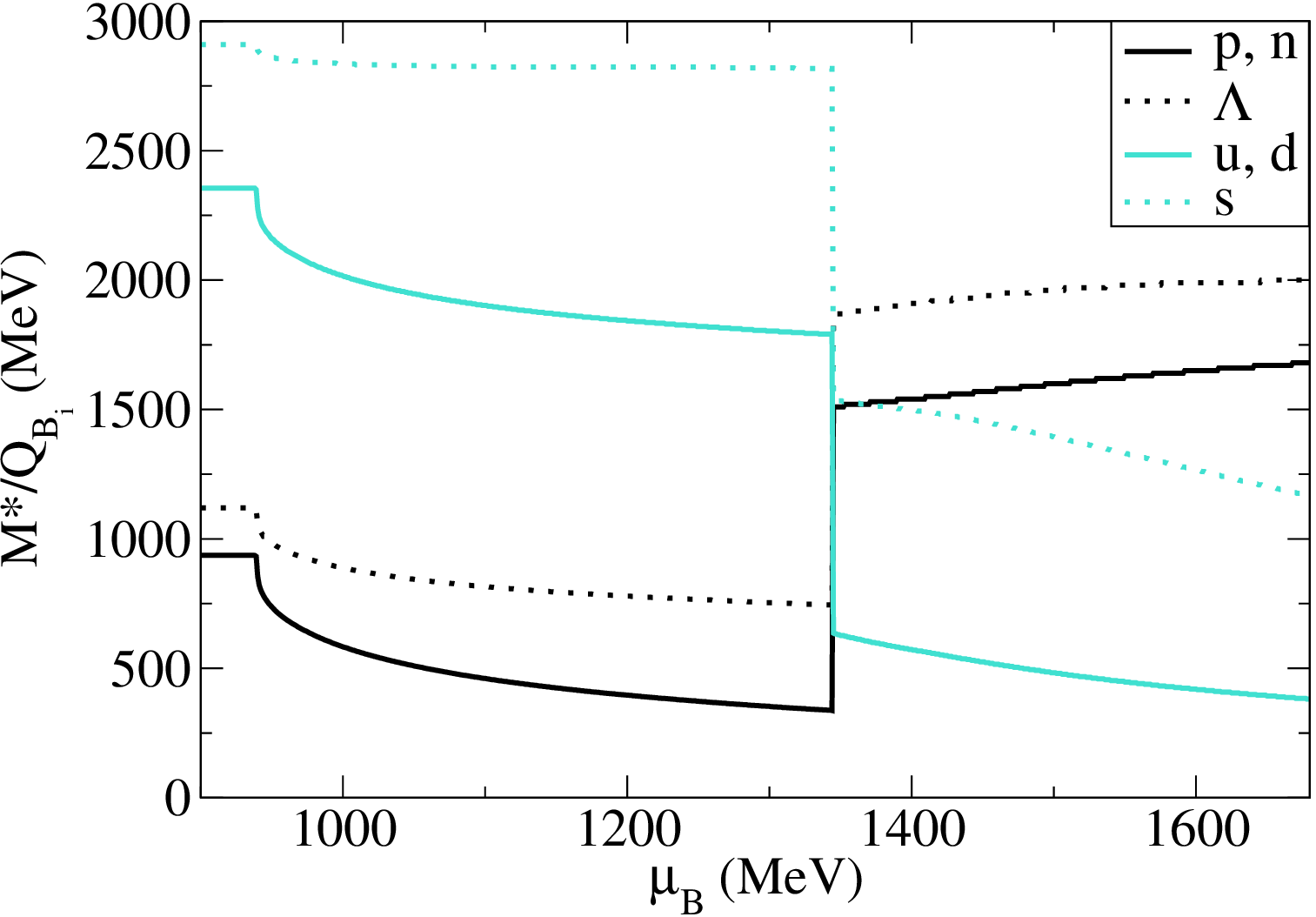}
\caption{\label{Meff}(Color online) Effective normalized mass of  different
species as a function of baryonic chemical potential for star matter at zero
temperature \cite{Dexheimer2010}.}
\end{figure}

The effective masses of the baryons and quarks are given by
\begin{eqnarray}
&M_{B}^*=g_{B\sigma}\sigma+g_{B\delta}\tau_3\delta+g_{B\zeta}\zeta+M_{0_B}+g_{
B\Phi} \Phi^2,&
\label{6}
\end{eqnarray}
\begin{eqnarray}
&M_{q}^*=g_{q\sigma}\sigma+g_{q\delta}\tau_3\delta+g_{q\zeta}\zeta+M_{0_q}+g_{
q\Phi}(1-\Phi),\nonumber&\\&
\label{7}
\end{eqnarray}
where $M_0$ is equal to $150$ MeV for nucleons, $354$ MeV for hyperons, $5$
MeV for up and down quarks and $150$ MeV for strange quarks.

Equations (\ref{6}) and (\ref{7}) show that as the field $\Phi$
increases (with the increase in density/temperature) baryons are
supressed, giving way to the quark phase, effectively modeling the QCD deconfinement phase
transition. The opposite is true for low values
of $\Phi$ (at low density/temperature).

The effective normalized masses of baryons and quarks are shown in Fig.
\ref{Meff}.
  Since the coupling constants in the $\Phi$ term of the effective mass formulas
are high but still finite,  the effective masses of the species not present
in each phases are large but also finite.

In analogy to the PNJL model, we define the potential $U$ for
$\Phi$ as
\begin{eqnarray}
&U=(a_0T^4+a_1\mu^4+a_2T^2\mu^2)\Phi^2&\nonumber\\&+a_3T_0^4\log{(1-6\Phi^2+8\Phi^3-3\Phi^4)}.&
\end{eqnarray}

In our case, $U(\Phi)$ is a simplified version of the potential used in
\cite{Ratti2006,Roßner2007} and adapted to  include terms that
depend on the chemical potential. These two extra terms
are not unique, but the most
simple natural choice of extending the potential. The corresponding
parameters are chosen to reproduce
the main features of the phase diagram at finite densities.
The coupling constants for the quarks can be found in reference
\cite{Dexheimer2010},
and are chosen to reproduce lattice data as well as known information about the
phase diagram. The lattice data includes a first order phase transition at
$T=270$ MeV, and a pressure
function P(T) similar to refs \cite{Ratti2006,Roßner2007} at $\mu=0$ for pure
gauge (for the quenched case without hadrons and quarks). 

\begin{figure}
\centering
\vspace{1.0cm}
\includegraphics[width=8.5cm, height =  6.0cm]{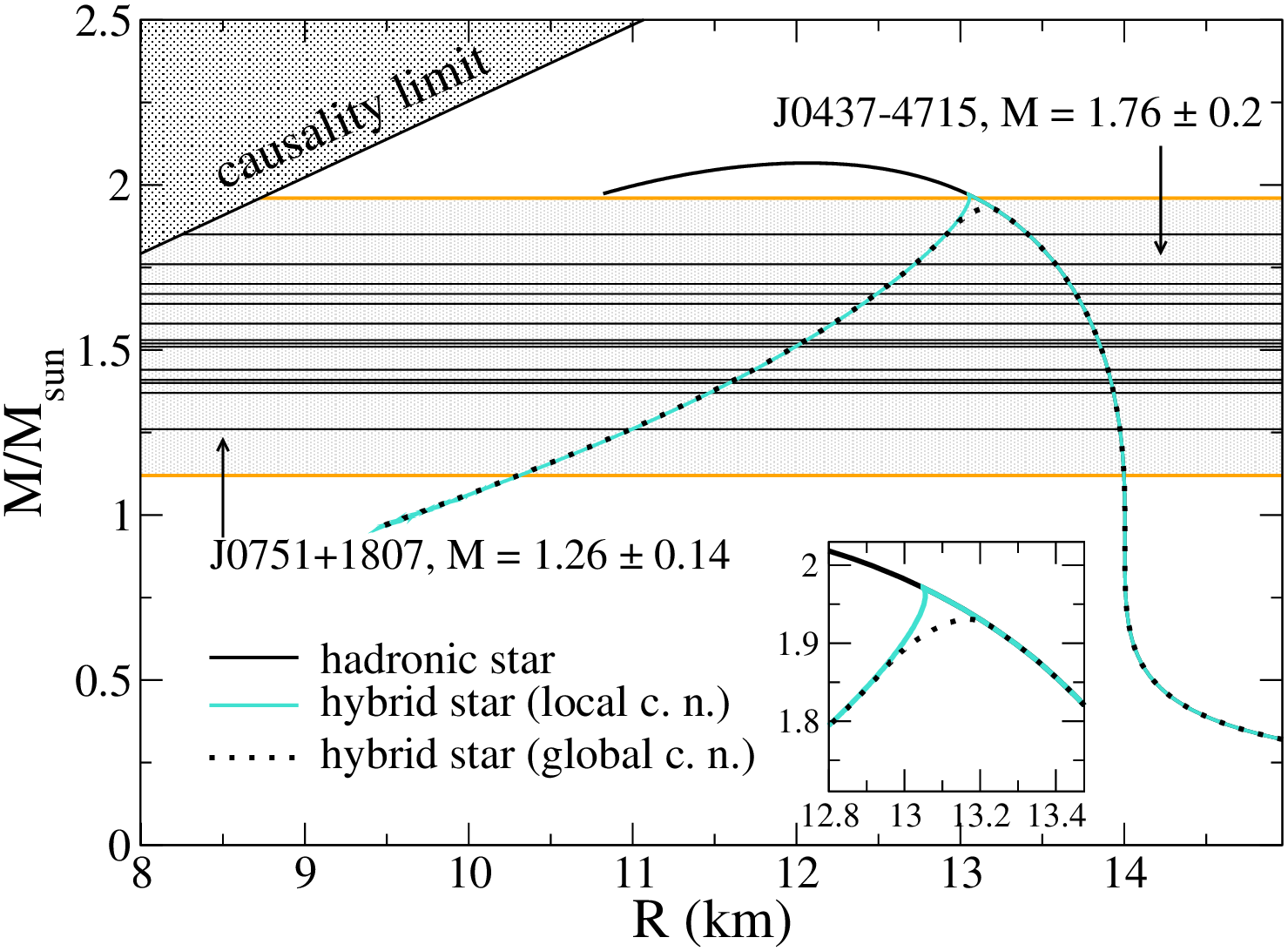}
\caption{\label{mass}(Color online) Mass-radius diagram for the model
investigated in this paper. The horizontal lines represent
observed pulsar masses
(\cite{Manchester2005,Stairs2004,LATTIMER2007,Verbiest2008,Freire2008,
Champion2008,Kurkela2010} and references therein) . The band delimited
by the two thick horizontal
lines (orange lines online) represent the range of pulsars masses observed,
accounting for the error of the highest and lowest observed mass.
}
\end{figure}

\section{Results}
\subsection{Deconfinement to Quark Matter}

\begin{figure}
\centering
\vspace{1.0cm}
\includegraphics[width=8.45cm]{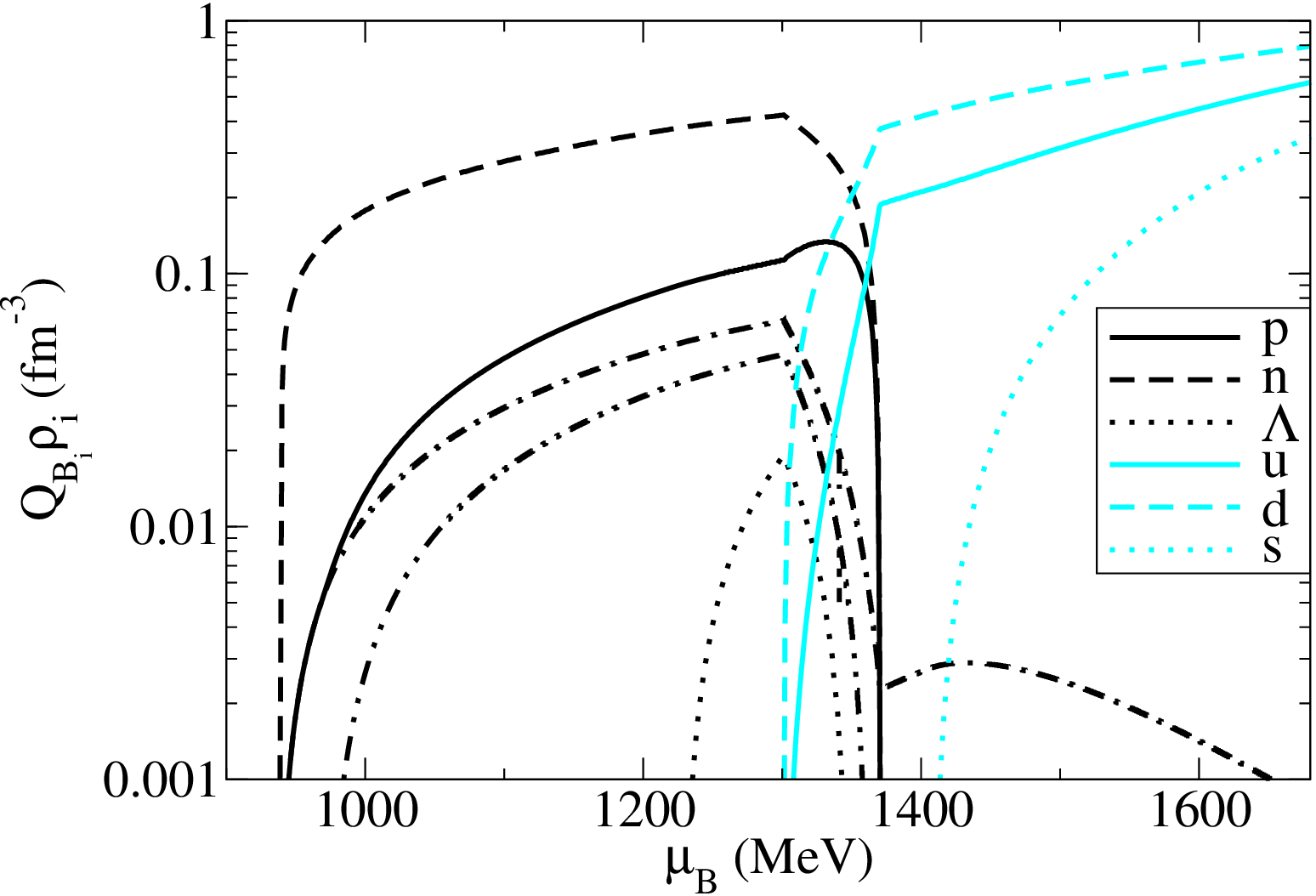}
\caption{\label{popgib}(Color online) Population  (baryonic density for
different species as a function of baryonic chemical potential) for star matter
at zero temperature using global charge neutrality \cite{Dexheimer2010}.}
\end{figure}

In our model the quarks are 
suppressed in the hadronic phase and the hadrons are
suppressed  in the quark phase, up to  $\mu_B = 1700$ MeV for  $T=0$.
This behavior is due to the fact that the coupling constants in the $\Phi$ term
of the effective mass formulas are high but still finite, so in principle at very high
chemical potential the threshold can be reached a second
time for hadrons. This threshold, which is higher than the density in the center
of neutron stars, establishes a limit for the applicability of the model.
The hyperons, despite being included in the calculation, are suppressed by
the appearance of the quark phase. Only a very small amount of Lambdas
appears immediately before the phase transition.
The strange quarks appear after the other quarks with relatively low abundance.
The density of electrons and muons is significant in the hadronic phase but not in the quark
phase, since the down and strange quarks are negatively charged, reducing
the need for leptons to maintain charge neutrality.

\begin{figure}[h]
\centering
\vspace{1.0cm}
\includegraphics[width=8.45cm]{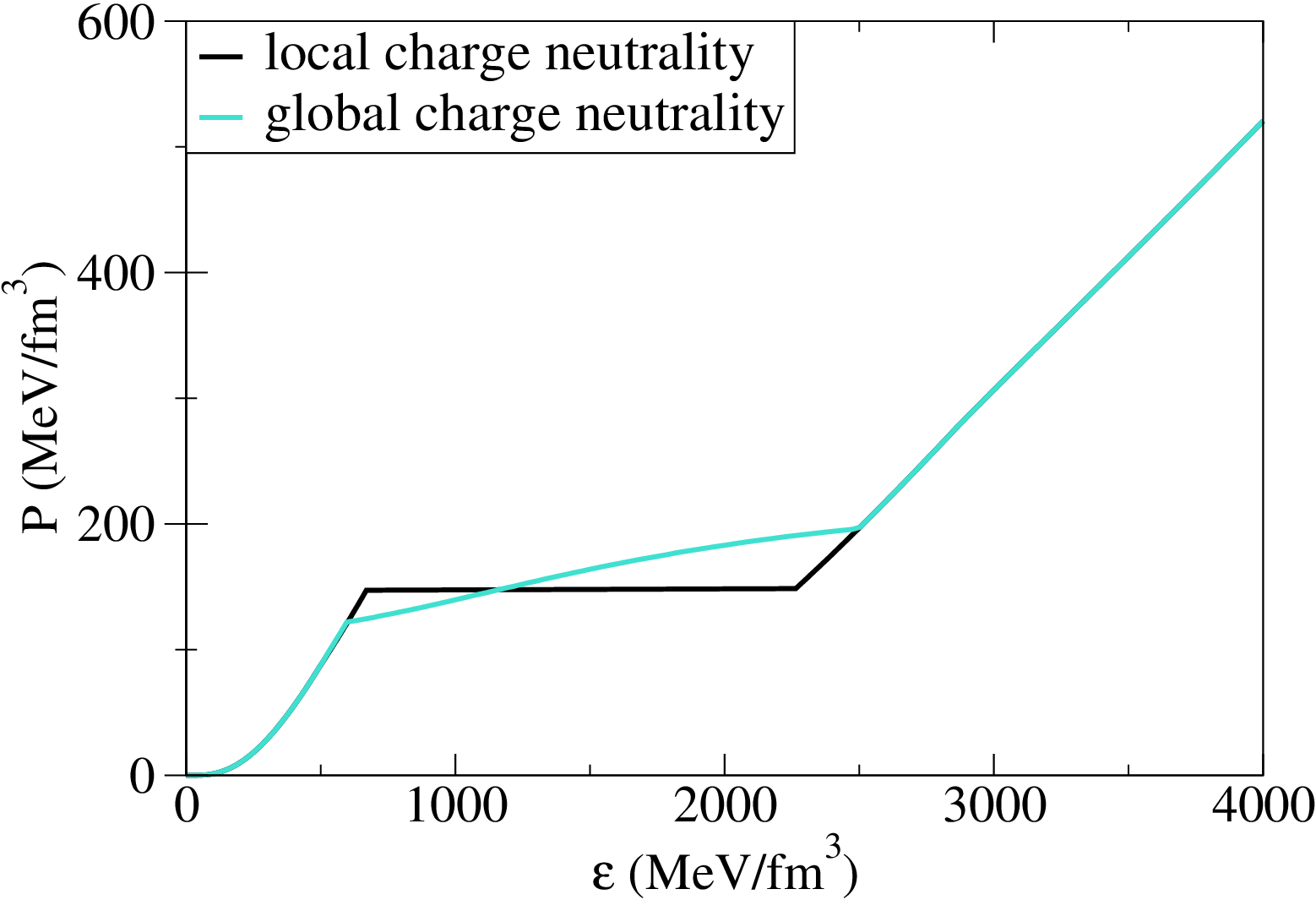}
\caption{\label{eos}(Color online) Equation of state (pressure as a function of
energy density) for star matter at zero temperature using local and global
charge neutrality \cite{Dexheimer2010}.}
\end{figure}

Focusing on the high-density/low-temperature region of the phase diagram we are able to calculate neutron star masses and radii
solving the Tolmann-Oppenheimer-Volkof equations
\cite{Tolman1939,Oppenheimer1939}.
The solutions for hadronic (same model but without quarks) and hybrid stars are
shown in Fig.~\ref{mass}, where
besides our equation of state for the core, a separate equation of state was used for the
crust \cite{Baym1971}. The horizontal lines in Fig.~\ref{mass} represent
observed masses of some prominent millisecond pulsars
(\cite{Manchester2005,Stairs2004,LATTIMER2007,Verbiest2008,Freire2008,
Champion2008,Kurkela2010} and references therein). To avoid cluttering
the graph we only indicated the name of the pulsars that establishes the lowest
and highest observed masses: J0751+1807 with mass $M/M_\text{sun} = 1.26 \pm
0.14$ \cite{Freire2008} and J0437-4715 with $M/M_\text{sun} =
1.76 \pm 0.2$ \cite{Manchester2005,Verbiest2008} respectively. Any plausible
equation of state must be able to produce neutron stars within the range
delimited
by these two objects, as is the case for the model investigated in this paper.
Our model predicts a maximum neutron star mass of $2.1~M_\odot$ considering
local charge neutrality, which as shown in Fig.
~\ref{mass}, is in agreement with the observational  constraints shown. Comparing to the
hadronic star sequence, the hybrid sequence has a much sharper peak. This peak (which
denotes the most massive stable star in the sequence) signals the phase transition into
quark matter at the core of the star.   Because the
equation of state for quark matter is much softer than the one for hadronic
matter, the star becomes unstable at the point where the central density is
higher than the phase transition threshold.

\begin{figure}[h]
\centering
\vspace{1.0cm}
\includegraphics[width=8.45cm]{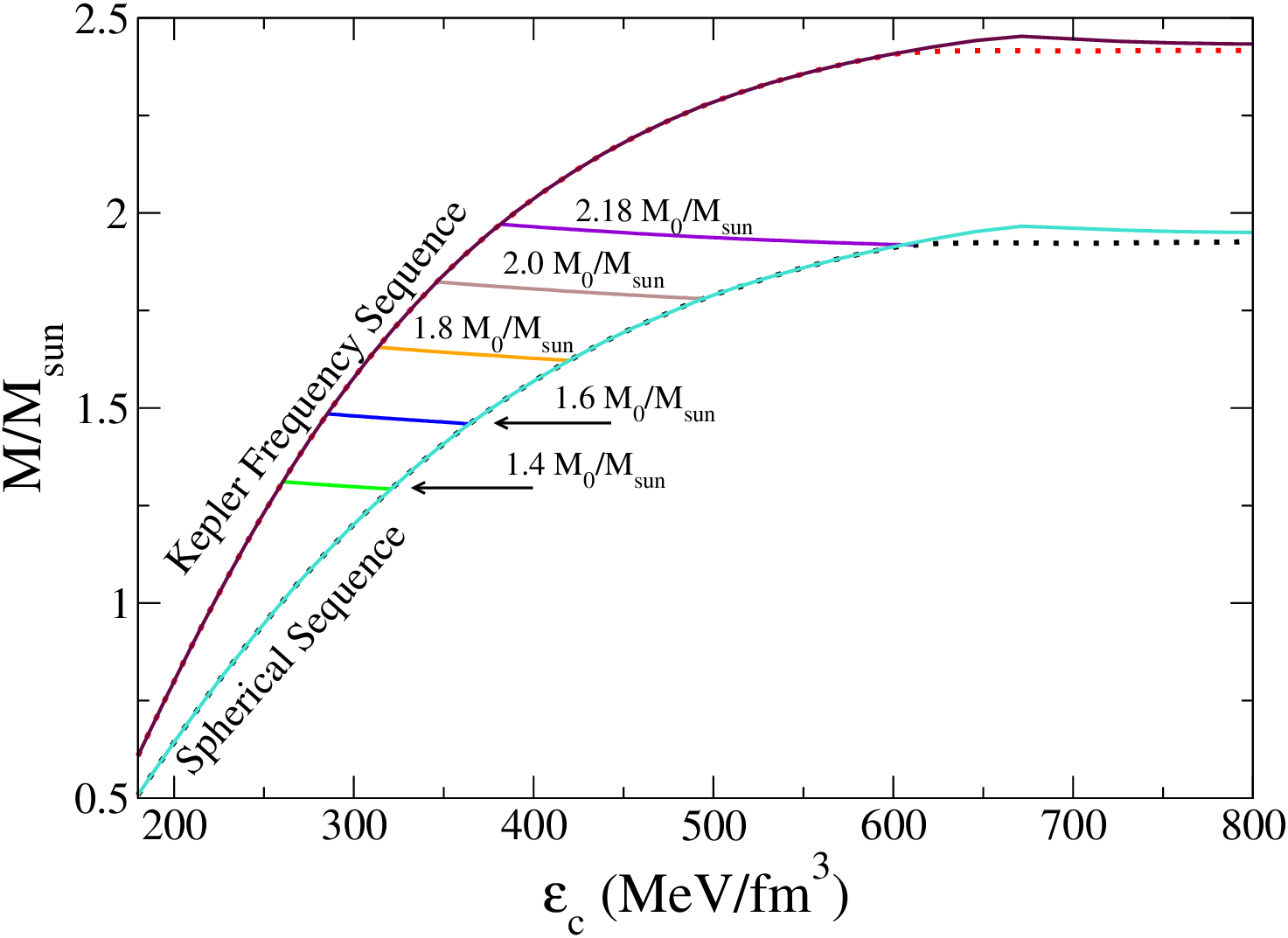}
\caption{\label{Massxec}(Color online) Gravitational mass as a function of
central density. The lowermost sequences represent static stars ($\Omega = 0$
Hz), whereas the higher most represents stars rotating at their Kepler
frequency.
The curves connecting the Kepler frequency to the static sequence are for stars
with constant baryonic mass. These sequences represent the evolution of Kepler
frequency stars to non-rotating objects.}
\end{figure}

There is still another possible option for the configuration of the particles in the neutron
star \cite{Glendenning1992}. If instead of local we consider global charge
neutrality,
we find a mixture of phases even for zero temperature as discussed in \cite{Dexheimer2009a}. This possibility, which is a more
realistic approach, changes the particle
densities in the coexistence region making them appear and vanish in a
smoother way (Fig.~\ref{popgib}). Therefore, the maximum mass allowed for the
star is slightly lower in this case than in the previous one, as can be seen
from the dotted line in Fig.~\ref{mass}; however, this possibility allows
stable hybrid stars with a small amount of quarks. The mixed phase constitutes
the inner core of the star up to a radius of
approximately 2 km. The equation of state for both cases is shown in
Fig.~\ref{eos}. The large jump in the pressure for the local charge neutrality
case explains why the neutron stars become immediately unstable after the phase
transition in this configuration.

\subsection{Rotational Effects}

The results shown in Fig. \ref{mass} are for static neutron stars without
including
rotational effects. The rotational nature of pulsars (which in some cases
may be rotating with frequencies as high as 700 Hz) warrants the
investigation of the structure of rapidly rotating neutron stars, which is
considerably more complicated than that of static objects. In References
\cite{Dexheimer2008a,Schramm2003} the
rotational effects of the hadronic part of our model were investigated by means
of
the improved Hartle-Thorne pertubative method \cite{Glendenning1994}. Here
we will extend this research by performing exact rotational calculations and
by including the quark-phase described in previous sections.
The numerical method used for the solution of Einstein's field equations, and
for the stellar structure of rapidly rotating neutron stars is based on the KEH
method \cite{Komatsu1989}, which basically consists of expanding the metric
functions
in terms of Green's functions, which can be iteratively integrated,
allowing us to calculate the structure of the star. This method has
been expanded by several authors, and  details can be found in References
\cite{Cook1992,Stergioulas1995}.

\begin{figure}[h]
\centering
\vspace{1.0cm}
\includegraphics[width=8.45cm]{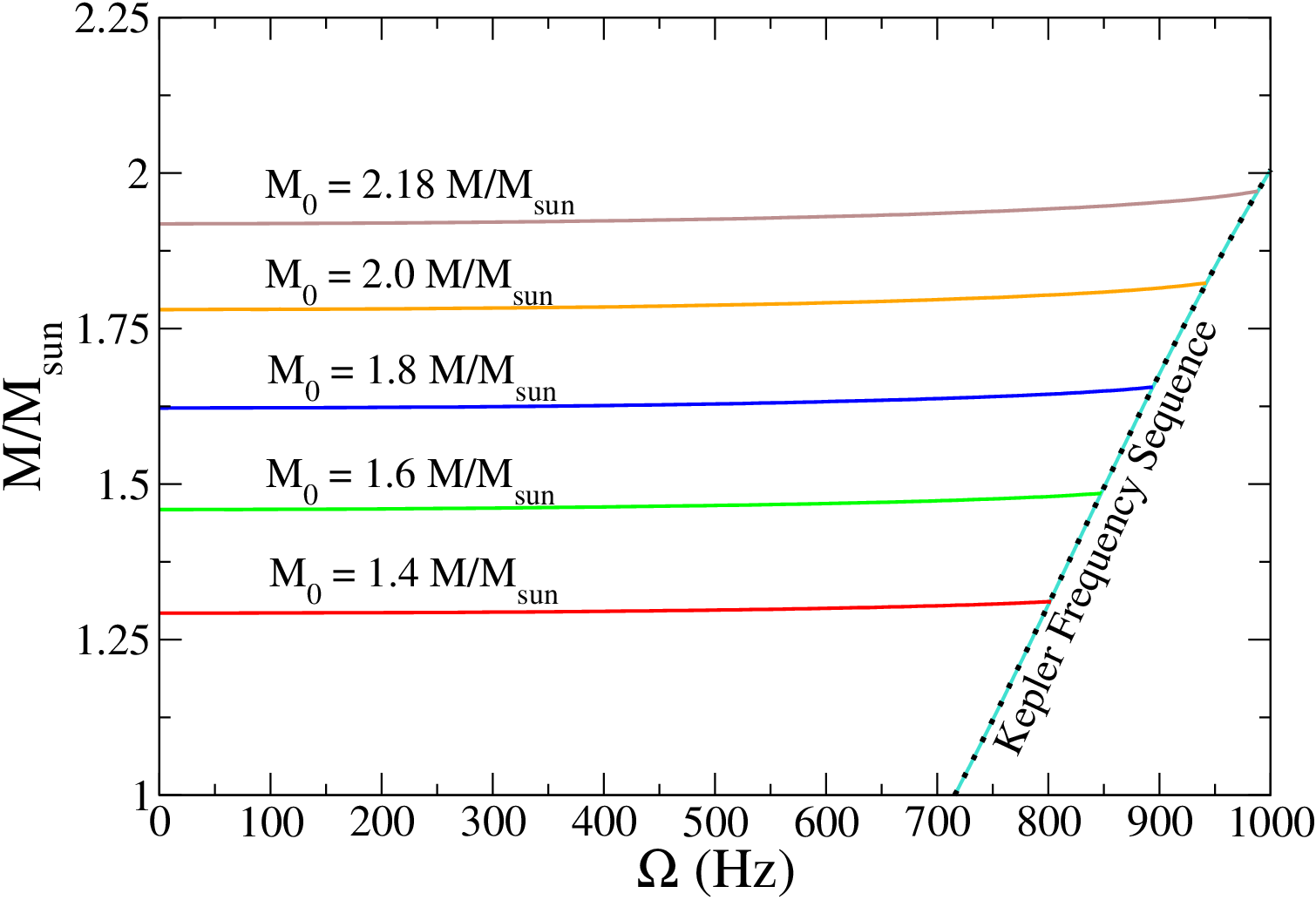}
\caption{\label{Massxfreq}(Color online) Gravitational mass as a function of
 frequency for the stars of Fig. \ref{Massxec}. The stars from the static
sequence are on the y-axis of the graph.}
\end{figure}

In Fig. \ref{Massxec} we show the mass as a function of central density. The
lowermost sequence represents static stars ($\Omega = 0$
Hz), whereas the highest one represents stars rotating at their Kepler
frequency. The Kepler frequency, or mass shedding frequency, is the maximum
frequency at which a compact star may rotate. When rotating above this frequency
an object would shed mass at its equator. This quantity therefore sets an
absolute limit for the frequency of compact stars. The curves connecting the
Kepler frequency to the static sequence indicate stars with constant baryonic
mass.
These sequences represent the evolution of Kepler frequency stars to
non-rotating objects. This is better illustrated in Fig. \ref{Massxfreq},
which shows the gravitational mass as a function of frequency for the constant
baryon mass stars. Fig. \ref{Massxfreq} also shows that the gravitational
mass may be as high as 3\% higher for a star rotating at the Kepler frequency
(compared to a non-rotating star with same baryonic mass). The central
density, on the other hand, may be decreased by as much as 38\%, as shown in
Fig.\ref{Massxec}. It is important to stress that the constant baryon mass
sequences shown in Figs. \ref{Massxec} and \ref{Massxfreq} represents the
spin-down evolution of an isolated compact star (i.e. no accretion, hence the
constant baryon mass). Obviously the spin-down evolution is a function of time,
therefore the x-axis of Figs. \ref{Massxfreq} could just as
well be replaced by time. How the frequency (or equivalently the central
density) varies with time will depend on the spin-down rate of the star. The
computation of this quantity is not trivial, and depends on properties like
magnetic field and/or gravitational radiation emission. Those issues are beyond
the scope of the current paper, and therefore we show our results as a function
of frequency and central density, always keeping in mind the implicit time
dependence.

As shown in Figs. \ref{Massxec} and \ref{Massxfreq}, when rotating at their
Kepler frequency, neutron stars can attain higher masses. Loosely speaking 
this may be explained by the
centrifugal force that provides an extra support against gravitational
collapse. Even though we can find stable rotating neutron stars with masses up
to $\sim 2.5$ solar masses, these objects will collapse into black holes during
spin-down evolution, which might be observable by the sudden stop of the neutrino signal
originating from the star.
As shown in Fig. \ref{Massxec}, there is no stable
static neutron star with baryon mass greater than $2.18$ solar masses predicted by our
model.

\begin{figure}[h]
\centering
\vspace{1.0cm}
\includegraphics[width=8.45cm]{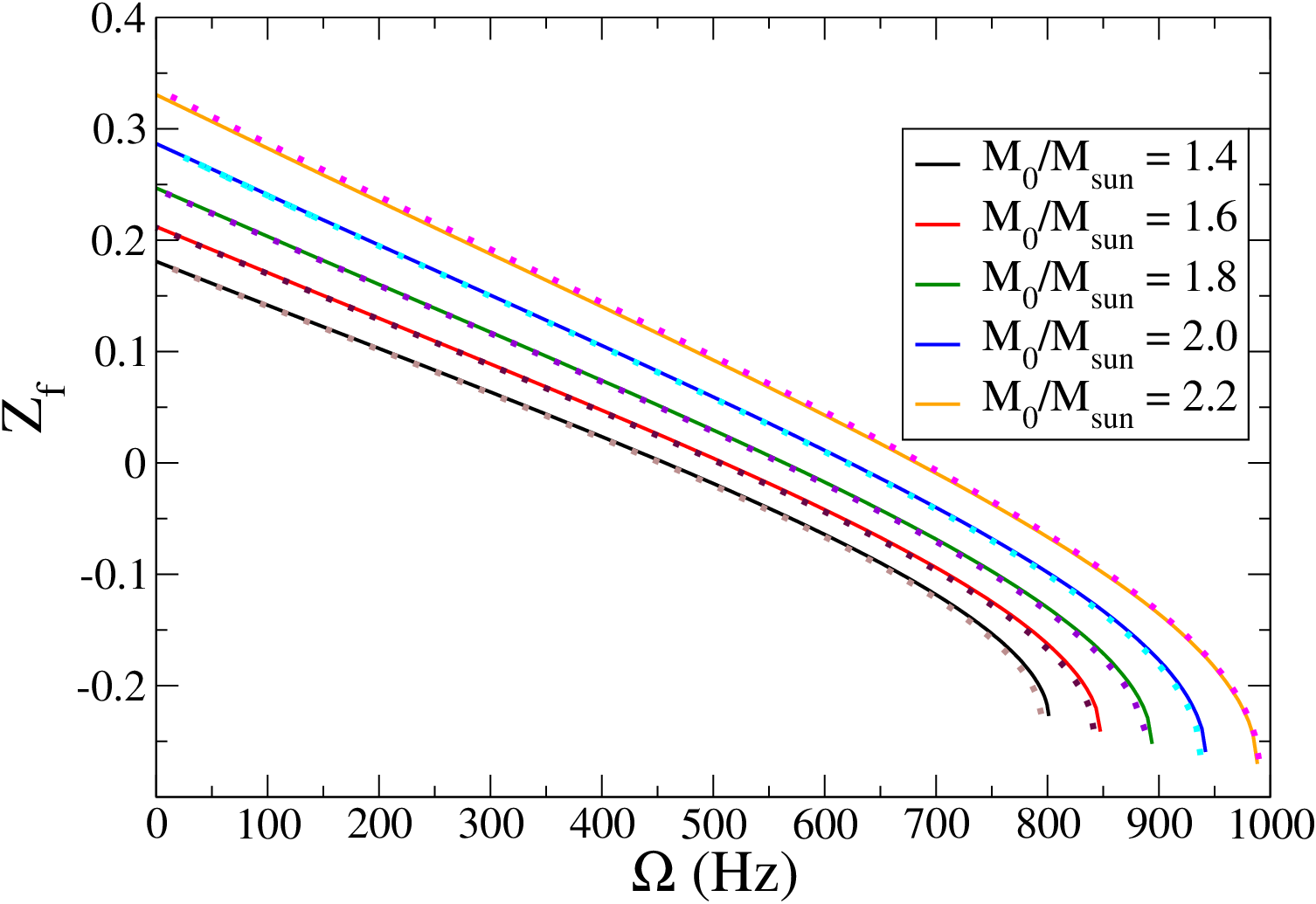}
\caption{\label{Zf_rot}(Color online) Forward redshift for the constant baryon
mass sequences of Fig. \ref{Massxec}. The dotted lines represent the redshift
 of the equivalent sequence for purely hadronic matter, which are almost
identical to that of Hybrid stars.}
\end{figure}

The rotation also alters the redshift of the stars significantly. In Fig.  
\ref{Zf_rot} we show the forward redshift as a function of frequency for the
stars of constant baryon mass of Fig. \ref{Massxec}. As one can see, the
forward redshift is substantially modified as the star's rotational frequency
is reduced. In the extreme case of high frequencies the forward redshift
becomes negative. For comparison purposes we have also plotted the redshift of
the equivalent stellar sequences with no quark phase (purely hadronic matter).
The redshift of these objects is almost identical to those of hybrid stars,
with a very slight deviation for very high frequencies. 


\begin{figure}[h]
\centering
\vspace{1.0cm}
\includegraphics[width=8.45cm]{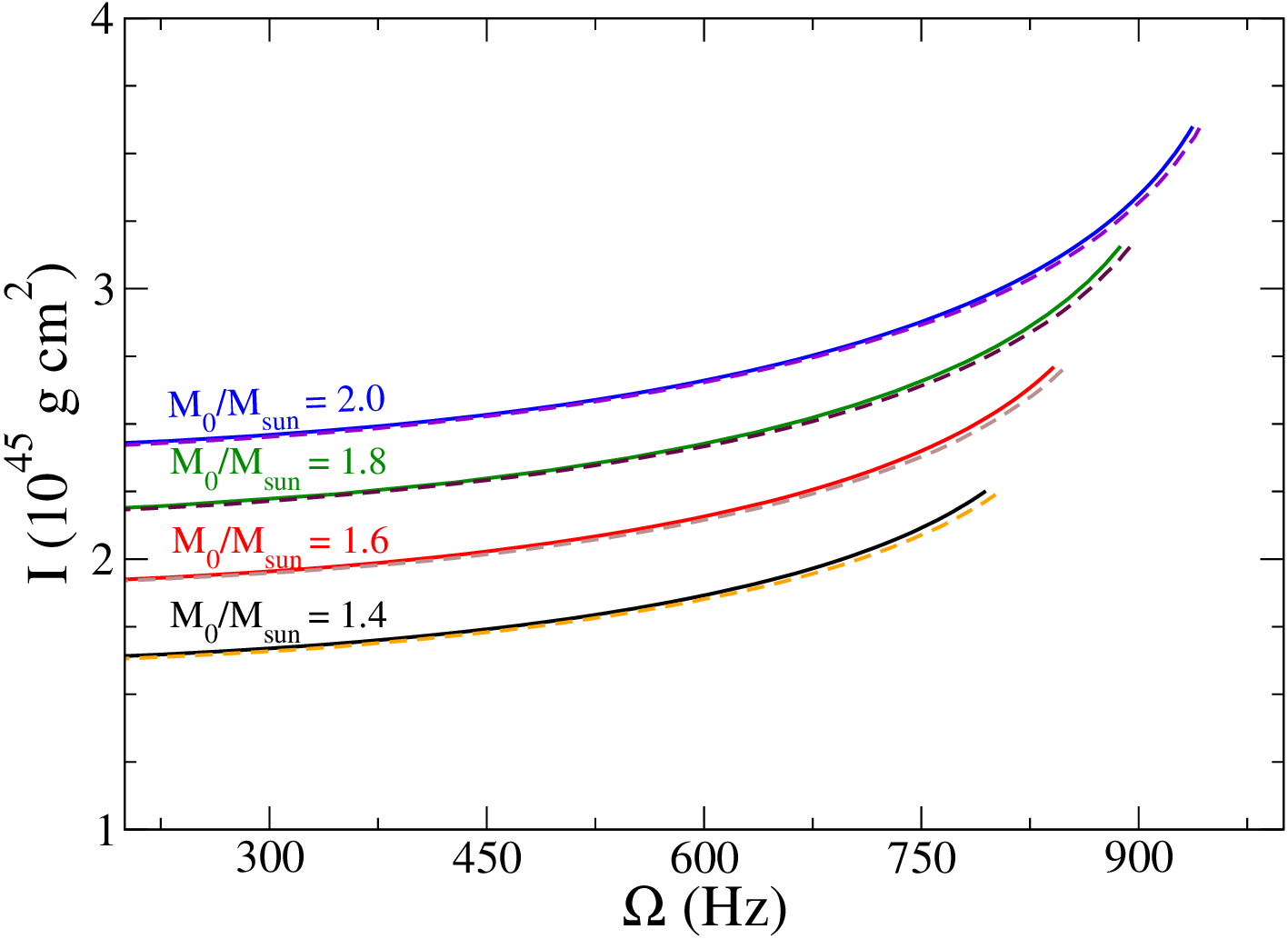}
\caption{\label{I_rot}(Color online) Moment of inertia for the constant Baryon
mass sequences of Fig. \ref{Massxec}. During the spin-down the geometry of
the objects change significantly, and therefore the moment of inertia for these
stars is substantially modified.}
\end{figure}

Just as the redshift is changed during the spin-down evolution, the moment of
inertia of the star should be modified as well. The moment of inertia as a
function of frequency for the stars shown in Fig. \ref{Massxec} are given in
Fig. \ref{I_rot}.

\begin{figure*}
\begin{center}
\begin{tabular}{cc}
\includegraphics[width=0.35 \textwidth]{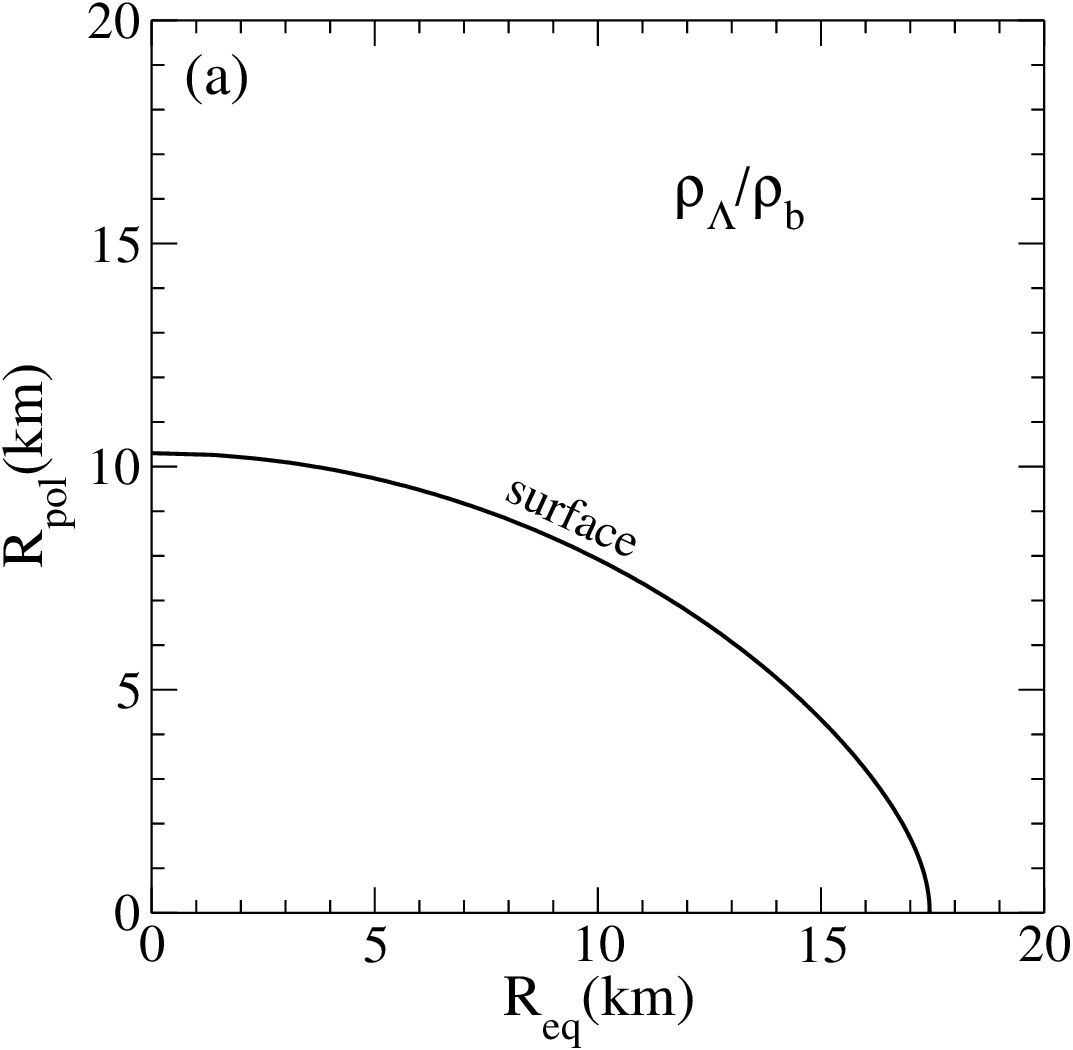} ~~&
\includegraphics[width=0.35 \textwidth]{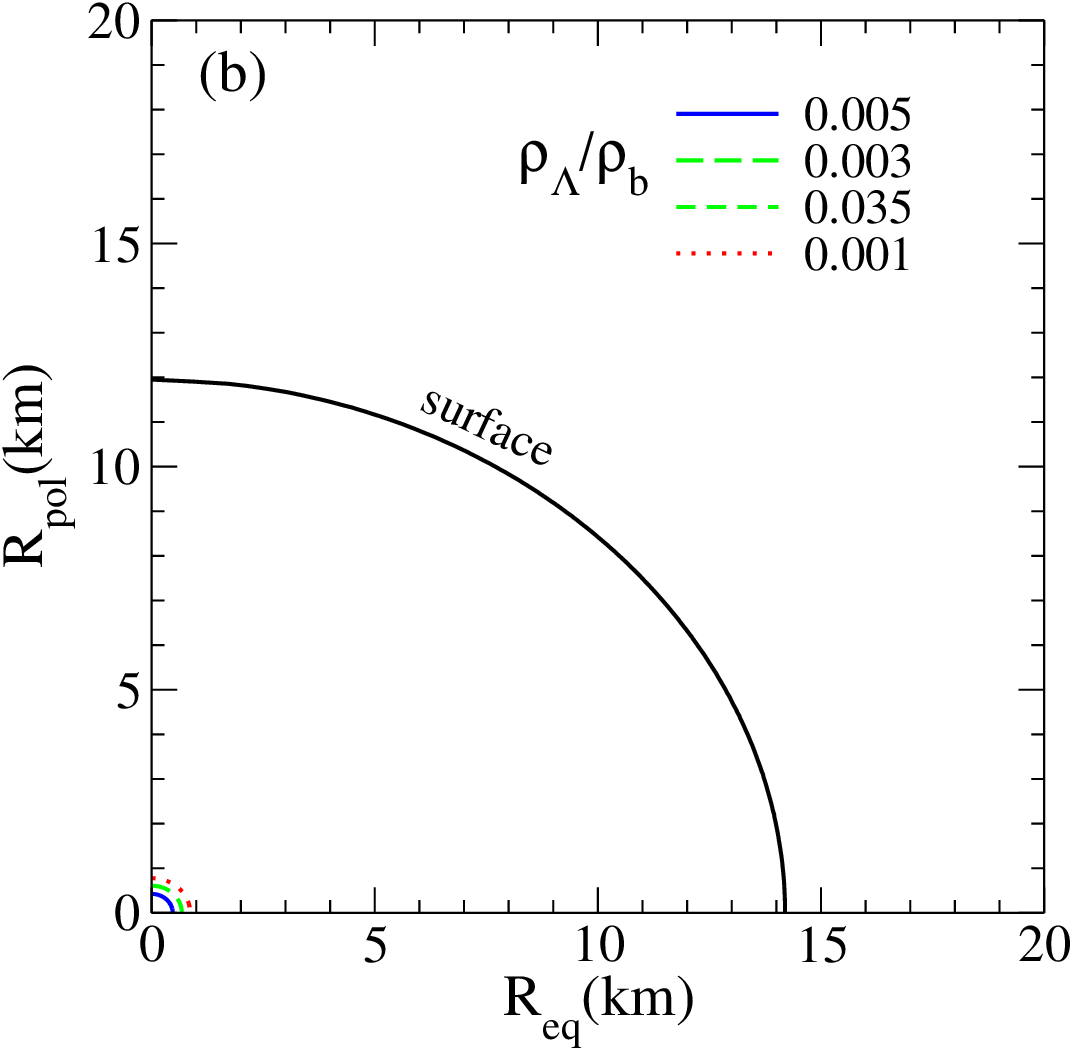} \\
 & \\
 & \\
 & \\
\includegraphics[width=0.35 \textwidth]{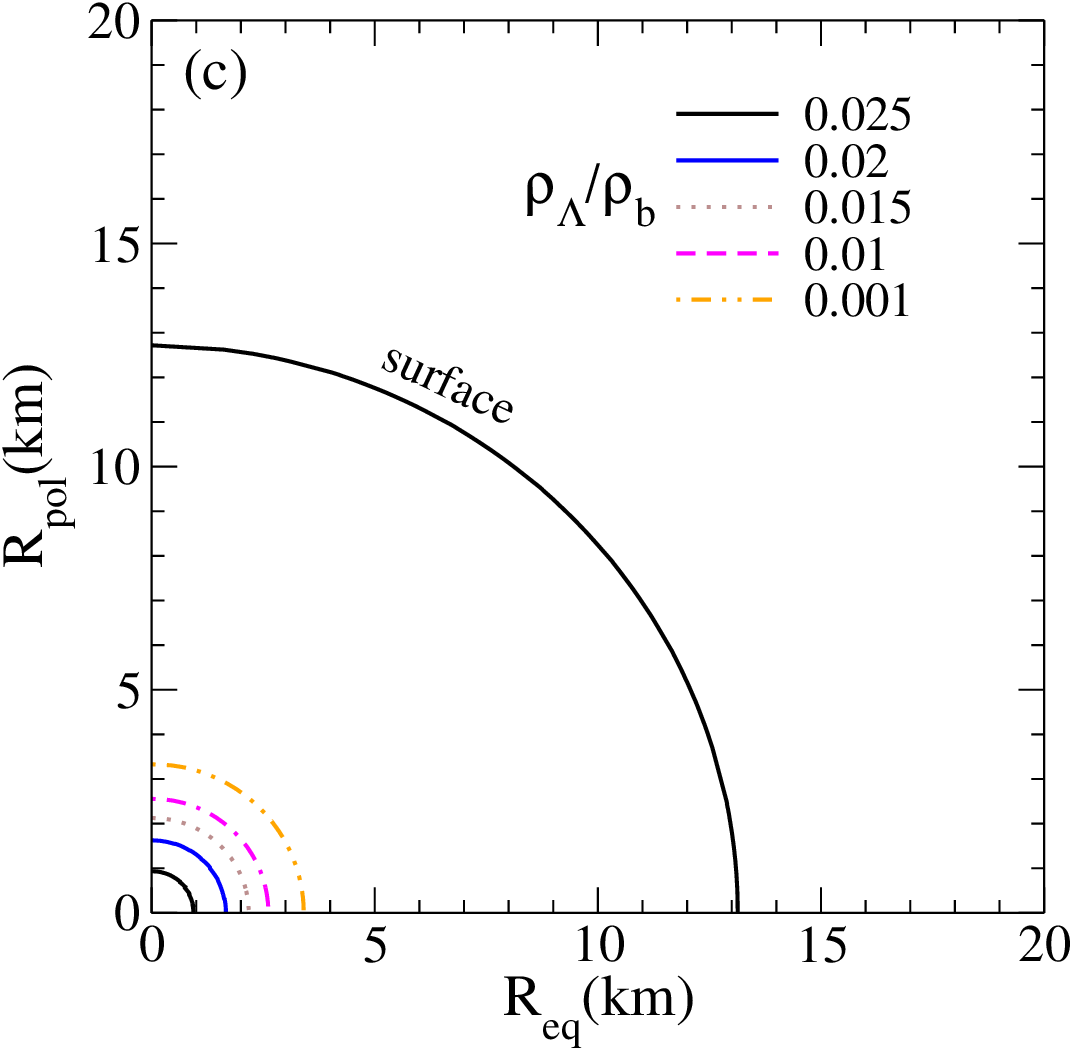} ~~&
\includegraphics[width=0.35 \textwidth]{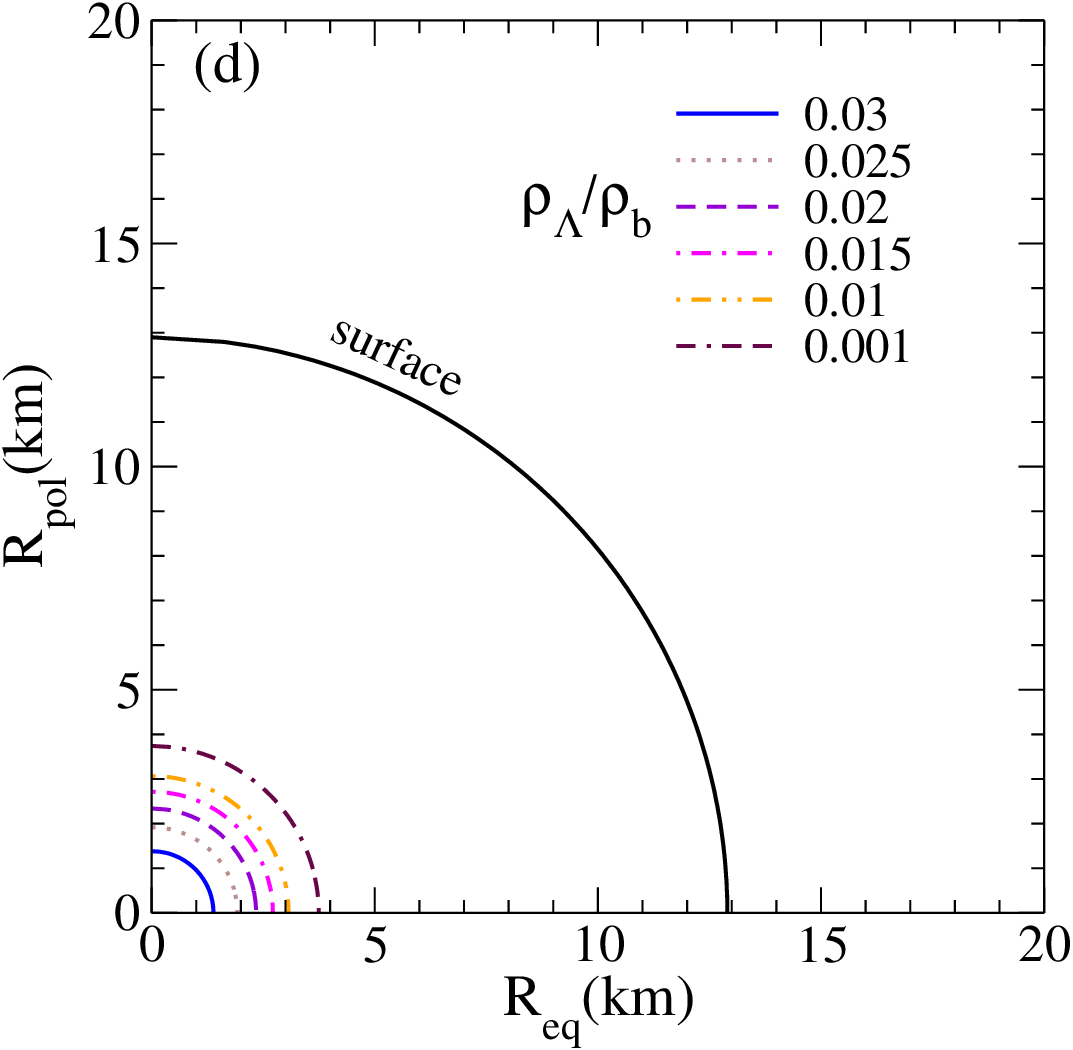}
\end{tabular}
\caption{ (Color online)Strangeness content for a compact star with
$M_0/M_{\text{sun}} = 2.18$, for frequencies of $988.4$ Hz (a), $746.0$ Hz (b),
$356.3$ Hz (c) and $0.0$ Hz (d). The z
axis represents
the relative population of $\Lambda$ ($\rho_\Lambda/\rho_b$).}
\label{Lamb_rot}
\end{center}
\end{figure*}

\begin{figure*}
\begin{center}
\begin{tabular}{cc}
\includegraphics[width=0.35\textwidth]{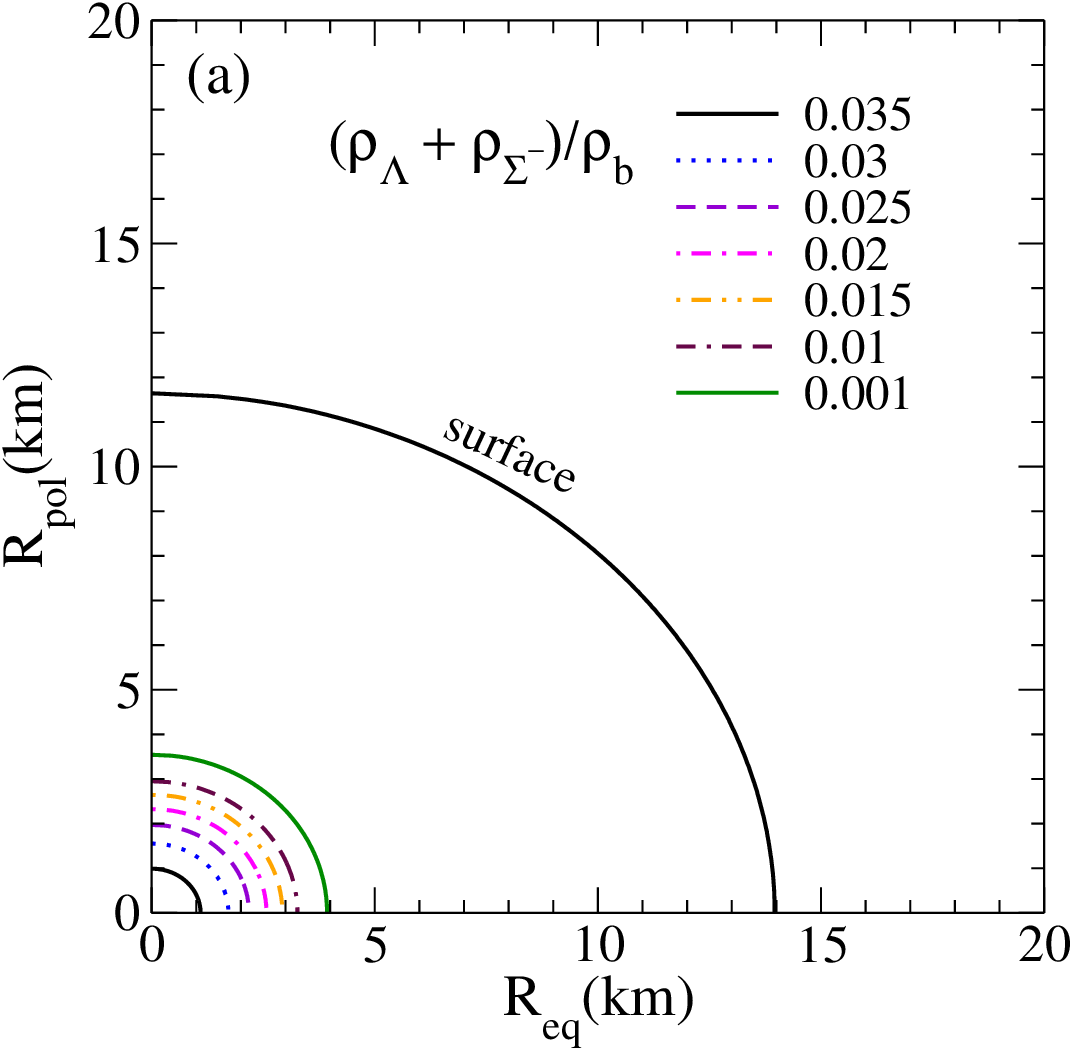} ~~&
\includegraphics[width=0.35\textwidth]{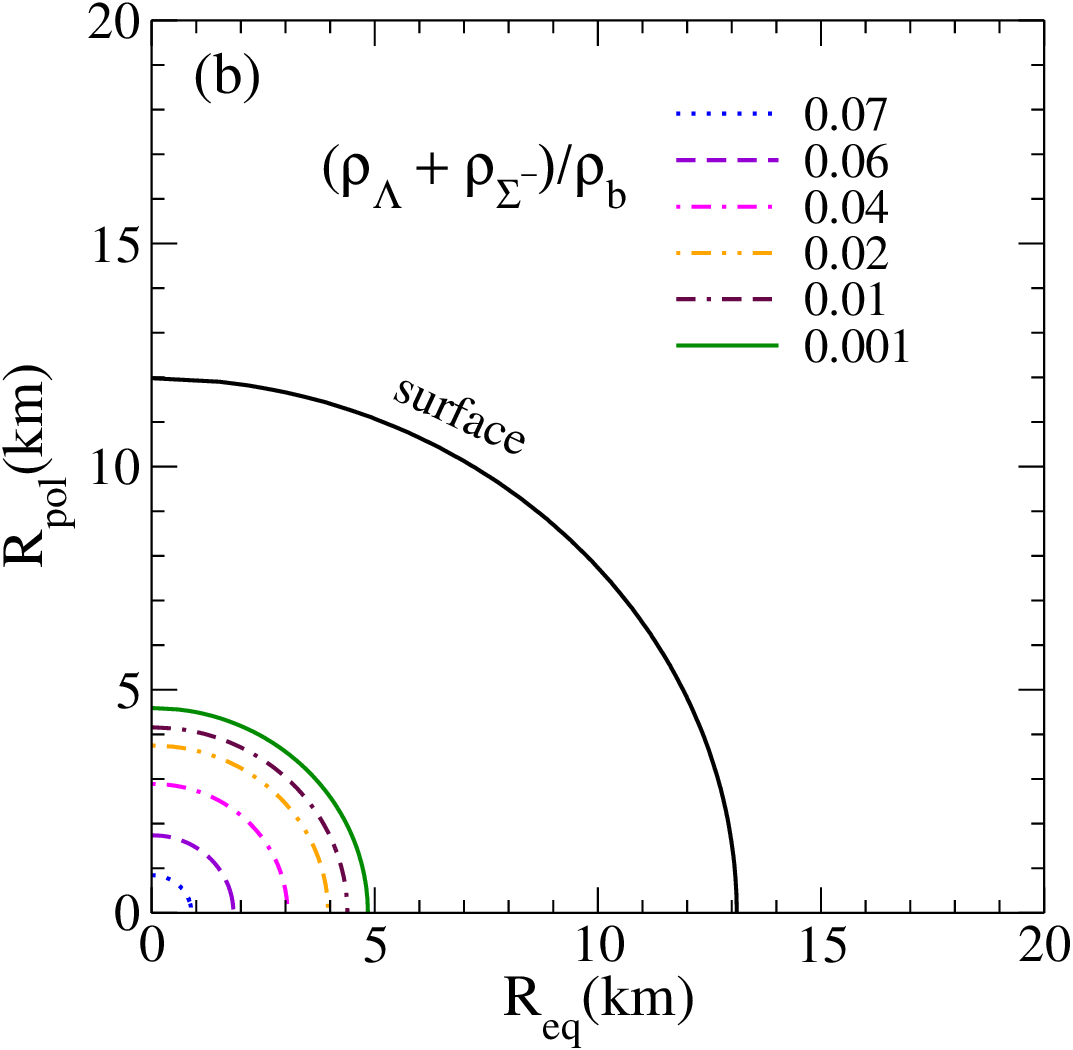} \\
& \\
 & \\
 & \\
\includegraphics[width=0.35\textwidth]{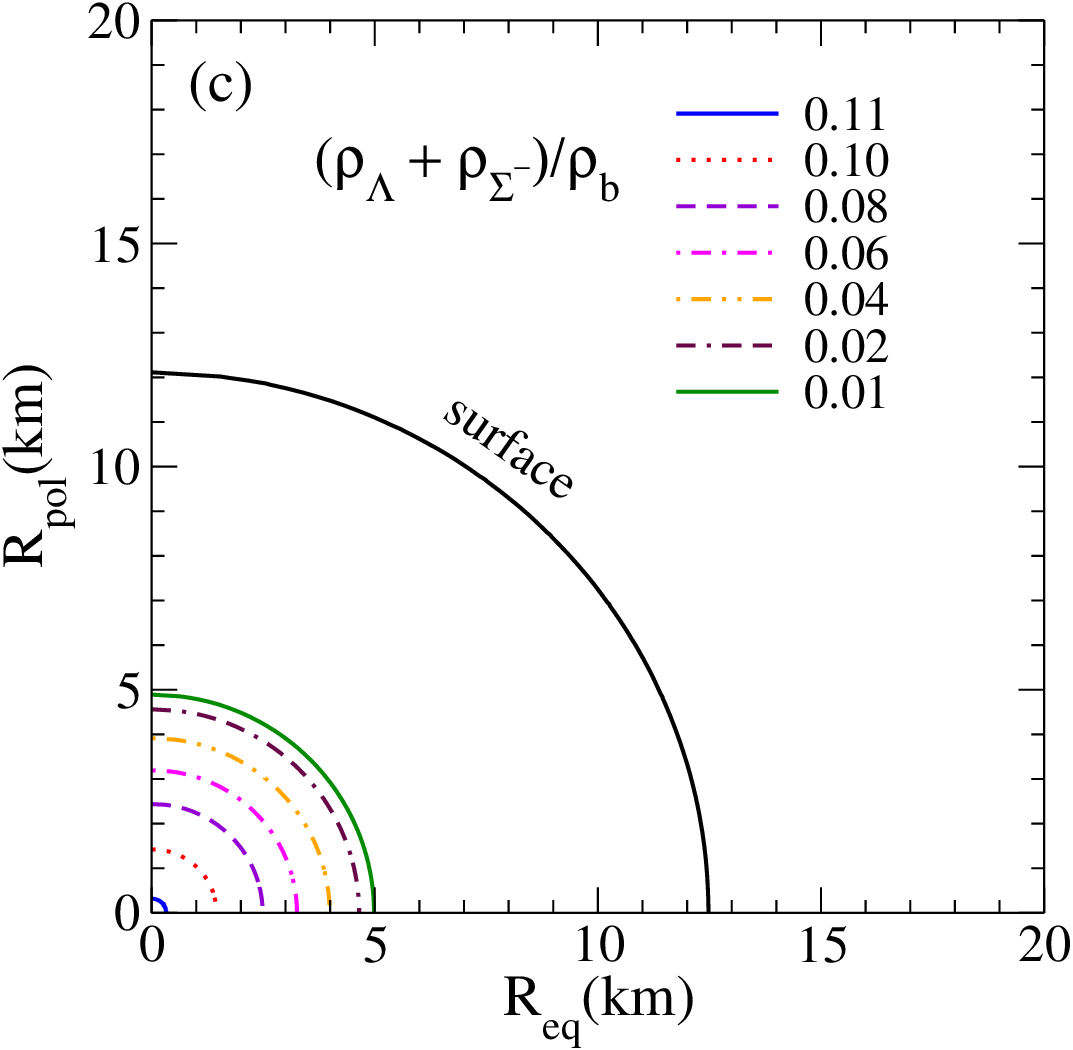} ~~&
\includegraphics[width=0.35\textwidth]{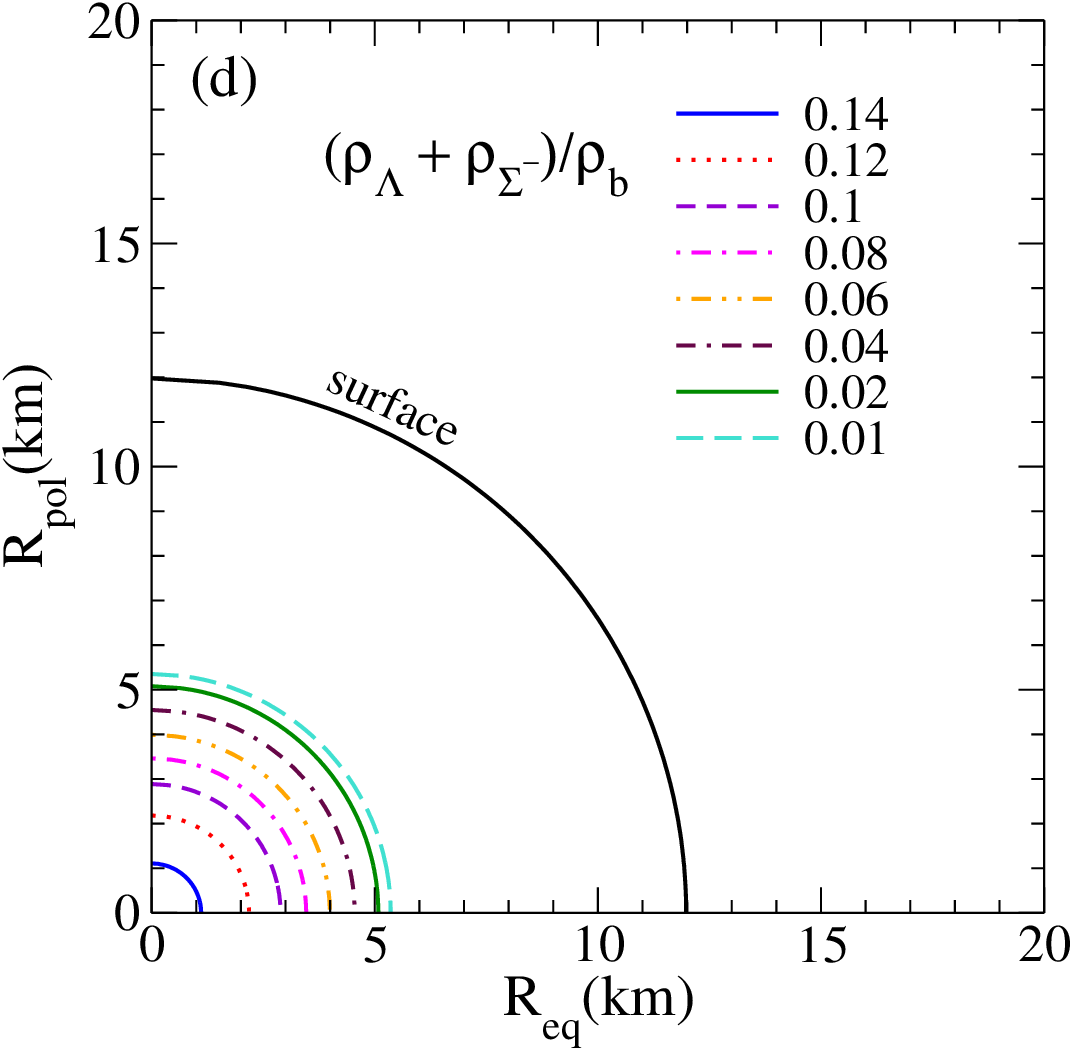}
\end{tabular}
\caption{(Color online)Strangeness content for a compact star with
$M_0/M_{\text{sun}} = 2.38$ in the pure hadronic model. The frequencies
are $828.3$ Hz (a), $639.9$ Hz (b), $394.8$ Hz (c) and $0.0$ Hz (d).  The z
axis represents
the relative population of $\Lambda$ and $\Sigma^-$ ($(\rho_\Lambda +
\rho_{\Sigma^-} /\rho_b$).}
\label{Lamb_rot_PH}
\end{center}
\end{figure*}

Another particular feature of the model investigated here is the relatively low level of
strangeness featured by the compact stars. As shown in Fig. \ref{popgib} the
$\Lambda$ states (the first strange particle states to be populated) start to
be occupied just before the onset of quark matter, which triggers the
unstable branch of compact stars. Thus, the phase transition to quark matter
suppresses the presence of strangeness in the object in the local charge
neutrality case. Only for the densest
objects we find a small region near the core with a low population of $\Lambda$
states. The spin-down will also affect the strangeness of the object,
increasing the radius of strangeness content near the core as the object
spins down and becomes denser. This result is shown in Fig. \ref{Lamb_rot}
where the relative population ($\rho_\Lambda/\rho_b$) for the $\Lambda$ states
is shown for
frequencies of $988.4$ Hz (a), $746.0$ Hz (b), $356.3$ Hz (c) and $0.0$ Hz (d).
This
result represents the strangeness content of the star, since no other
particles
containing strangeness are present. Fig. \ref{Lamb_rot}
shows that for higher frequencies (and thus lower densities) strangeness states
are either absent, or are very lowly populated. As the object spins down, and
its density increases, the strange core expands and becomes more highly
populated. For comparison purposes in Fig. \ref{Lamb_rot_PH} we also show the
strangeness content for a $M_0/M_{\text{sun}} = 2.38$  stars in the pure
hadronic model. In this case, since there is no phase transition to quark
matter, the stars are able to attain higher masses (as can be seen in Fig.
\ref{mass}). Furthermore the hyperons are not suppressed by the onset of the
quark phase, and these stars attain a higher strangeness content. As shown
in Fig. \ref{Lamb_rot_PH}, the $\Lambda$ fraction at the core of the
non-rotating star is 3.4 times higher than that found at the core of densest
(stable) hybrid star. Once more we recall that the different frequencies shown
in Figs. \ref{Lamb_rot} and \ref{Lamb_rot_PH} represent different stages of
time along the spin evolution of the star, with the time scale depending on the
spin-down rate of the object.

\begin{figure}[h]
\centering
\vspace{1.0cm}
\includegraphics[width=8.45cm]{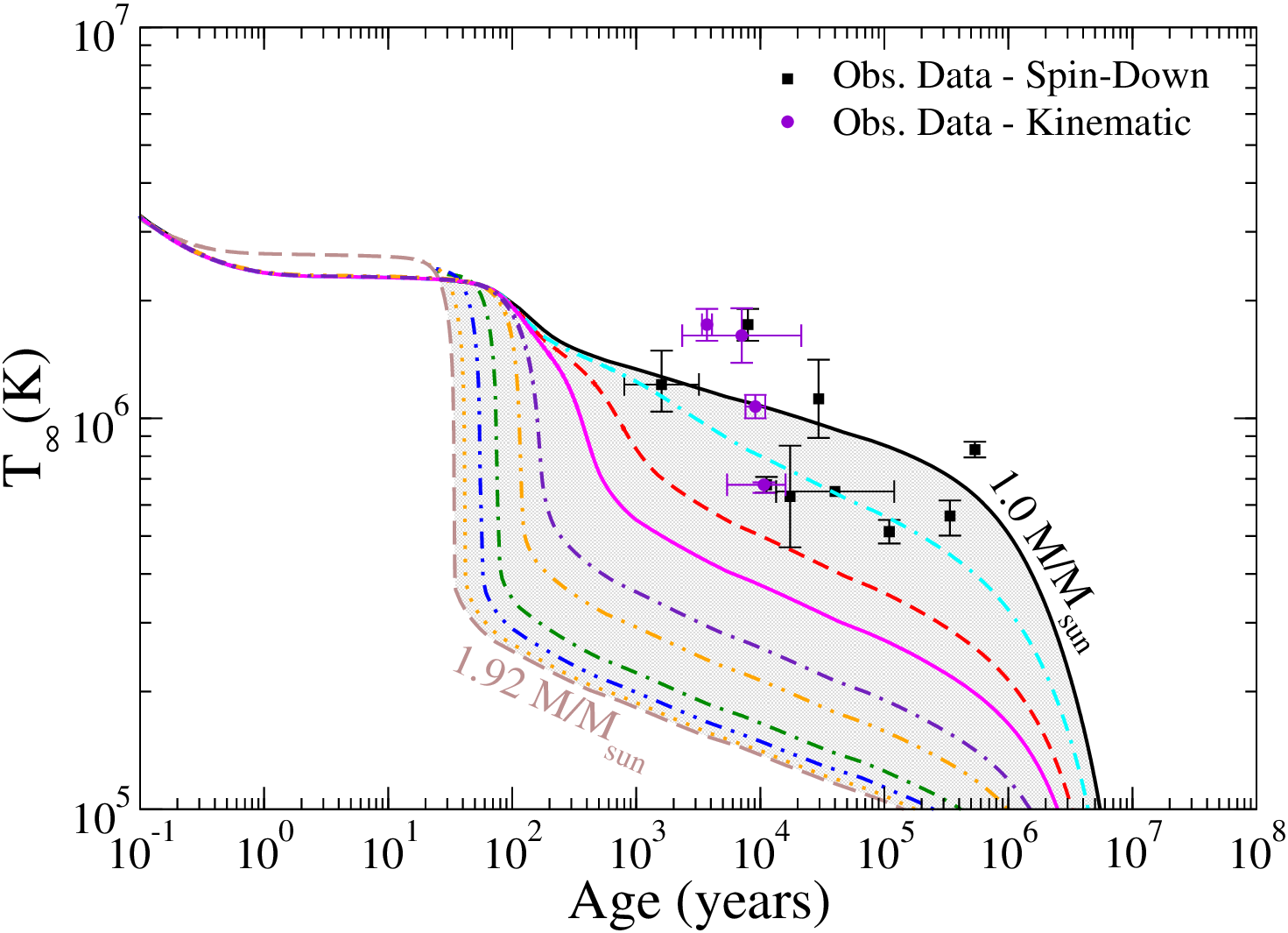}
\caption{\label{cooling}(Color online) Cooling of a range of spherically
symmetric neutron stars from Fig. \ref{mass}. The symbol $T_\infty$ denotes
the
redshifted temperature as detected by an observer at an infinite distance from
the star. The onset of the direct Urca
process happens for neutron stars with 1.22 solar masses. Any star with a lower
mass will have a slow cooling. For masses above this value an enhanced cooling
is achieved. Also shown is some prominent observed temperatures of neutron
stars \cite{Page2004,Page2009}. Squares denote that the age was estimated
based on the spin down rate, and circles indicate age estimates based on the
motion of the pulsar
with respect to its originating supernova remnant.}
\end{figure}

\subsection{Cooling Process}

Another method of probing the inner core of compact stars is by investigating
its thermal evolution. All the thermal processes taking place in a compact star
strongly depend on its composition, therefore by comparing theoretical
predictions with observed thermal data, one can obtain valuable information
regarding the core of neutron stars.

The thermal evolution of neutron stars is dominated by neutrino emissions for
the first 1000 years (maybe more in the slow cooling scenario) \cite{Page2006},
later being replaced by photon emission from the surface. The direct Urca
process \cite{Lattimer1991} (DU process hereafter) is the most efficient
cooling mechanism in a neutron star. With emissivities of the order of
$10^{26}$erg cm$^3$ s$^{-1}$ \cite{Lattimer1991}, neutron stars in which the DU
process takes place will cool very quickly. Due to momentum
conservation however, the DU process can only take place when the proton
fraction reaches a certain value (which depends on the underlying equation of
state, but it is usually between 11 -- 15 \%). Compact stars, whose proton
fraction are
below this threshold, will feature a slower cooling.
Other processes, like pairing and meson condensates, for example, might have an influence on
whether or not  a neutron star will feature a fast or slow cooling. These topics
are beyond the scope of the current research and will be pursued in
future investigations.

We solve the thermal evolution equations \cite{Weber} for the spherically
symmetric stars shown in Fig. \ref{mass}. The emission processes considered
for the core are the direct Urca process \cite{Lattimer1991}, the modified Urca
process \cite{Friman1979}, and the
Bremsstrahlung process \cite{Maxwell1979,Friman1979}; as for the crust we
consider the electron
Bremsstrahlung process \cite{Kaminker1999} , electron-positron annihilation
\cite{Yakovlev2001} and plasmon decay \cite{Yakovlev2001}  . The
results are shown in Fig. \ref{cooling}, where we show the cooling curves for
neutron stars with different gravitational masses. The symbol $T_\infty$ denotes
the
redshifted temperature as detected by an observer at an infinite distance from
the star.

We have also plotted some prominent observed temperatures of neutron stars
\cite{Page2004,Page2009}. Since there are two estimates for the age of
pulsars, we have plotted two sets of observed data. The first set is for
objects whose age estimate is based on the observed spin down rate, these are
represented by squares. The second set is for ages obtained by tracking the
pulsar back to its original supernova remnant (kinematic age, represented by
circles).

As shown in Fig. \ref{cooling} our model is in relatively good agreement with
the observed data, with the exception of a few high temperature pulsars. It is
possible that these
objects feature non-standard processes like pairing or some
re-heating mechanism which would explain their high temperature. The
shaded area represents all possible cooling curves for our model, that can be
obtained by solving the cooling equations for the whole spectrum of stable
hybrid stars shown in Fig. \ref{mass}.  We can also see that within our
model, neutron stars with masses higher than 1.22 solar masses, which feature
an enhanced cooling, cool down too fast to be in agreement with observed data.
It is important to mention that these results can be further improved if
one consider more sophisticated processes like pairing and meson condensation.

\begin{figure*}
\begin{center}
\begin{tabular}{cc}
\includegraphics[width=0.35\textwidth]{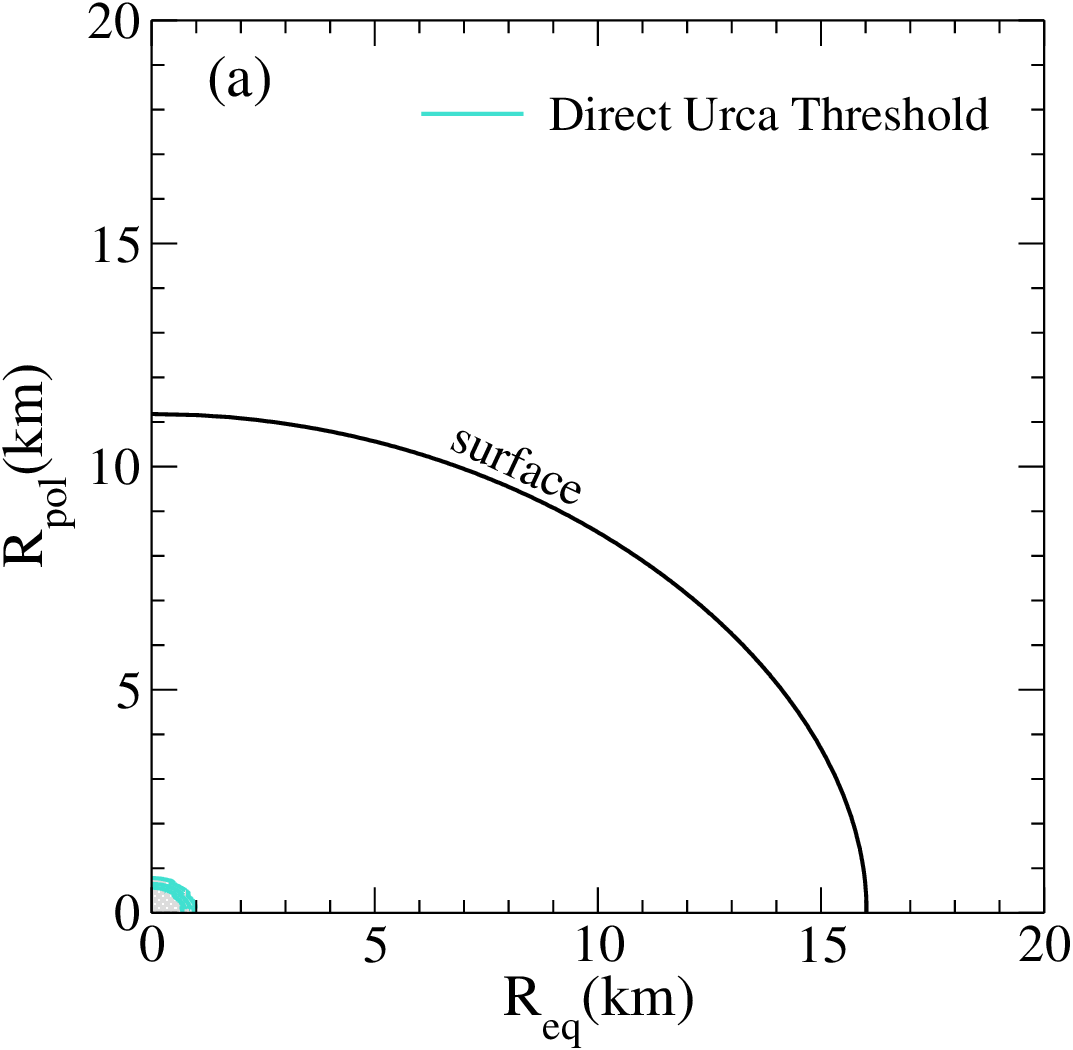} ~~&
\includegraphics[width=0.35\textwidth]{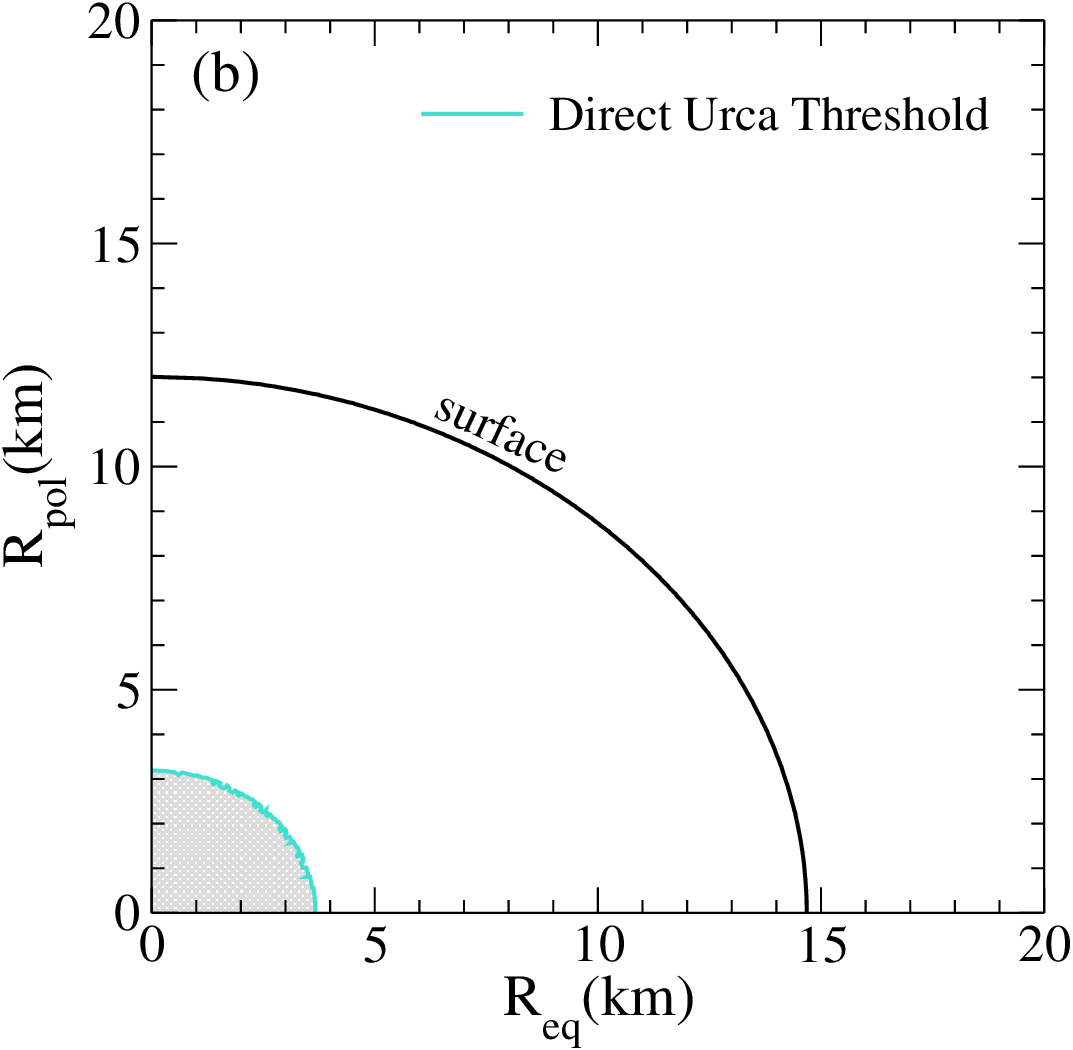} \\
& \\
 & \\
 & \\
\includegraphics[width=0.35\textwidth]{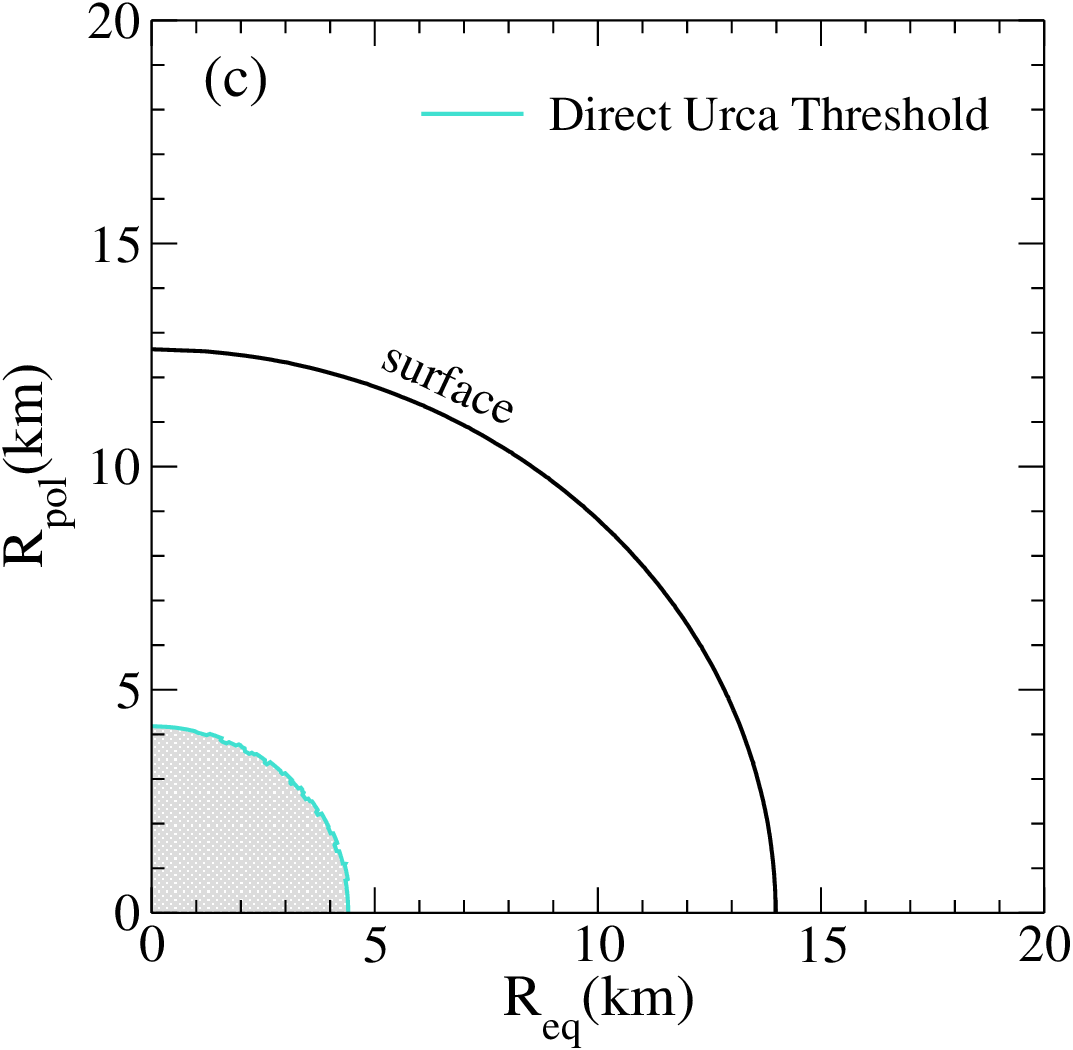} ~~&
\includegraphics[width=0.35\textwidth]{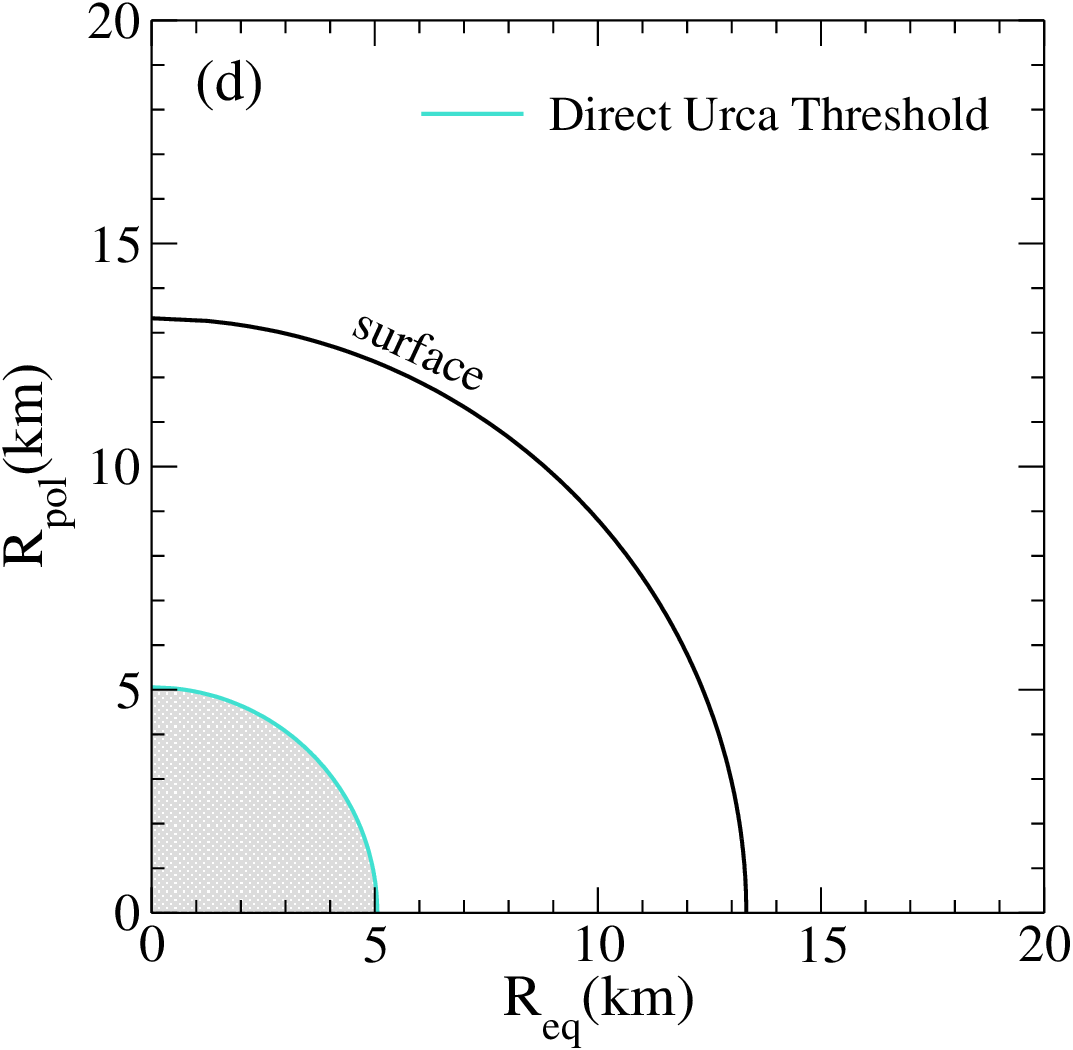}
\end{tabular}
\caption{(Color online) Direct Urca threshold for a compact star with
$M_0/M_{\text{sun}} = 1.6$, for frequencies of $792.4$ Hz (a), $656.8$ Hz (b),
$496.5$ Hz (c) and $0.0$ Hz (d). The grey shaded
areas
denotes the region in which the direct Urca process is allowed to take place
(denoted by z=1).}
\label{Urca_rad}
\end{center}
\end{figure*}

Finally we have also investigated how spin down effects might affect the
thermal evolution. As shown in Fig. \ref{Lamb_rot}, the increase in density
that follows spin down has a strong effect on the strangeness content of the
star. It is only natural to expect that other particle states will also be
altered by the spin down. Particularly interesting is the proton, electron and
neutron populations, since these particle population will dictate whether or not
the direct Urca process takes place. This process can only take place if the
following triangle inequality (and cyclic permutations of it) is satisfied
\begin{equation}
 k^f_n + k^f_p \geq k^f_e ,
\end{equation}
with $k^f_i$ denoting the Fermi momentum of particle $i$. We calculated the
radius threshold for different frequency stars, from the sequence with
$M_0/M_{\text{sun}} = 1.6$. This result is shown in Fig. \ref{Urca_rad},
where the gray shaded represents the region in which the DU process is
allowed to take place.

The results of Fig. \ref{Urca_rad} show that the region in which the DU
process takes place changes substantially during the evolution of the
compact star. For the case studied, for higher frequencies (where the density
is lower) the DU process can only take place in a very small region near the
core. This regions grows as the stellar frequency decreases and the object
becomes denser. This hints that the cooling of compact stars might be
slower if one consider effects of spin down. Current efforts are in progress to
better understand this issue, and will be discussed in a future paper.

\begin{figure}[h]
\centering
\vspace{1.0cm}
\includegraphics[width=8.45cm]{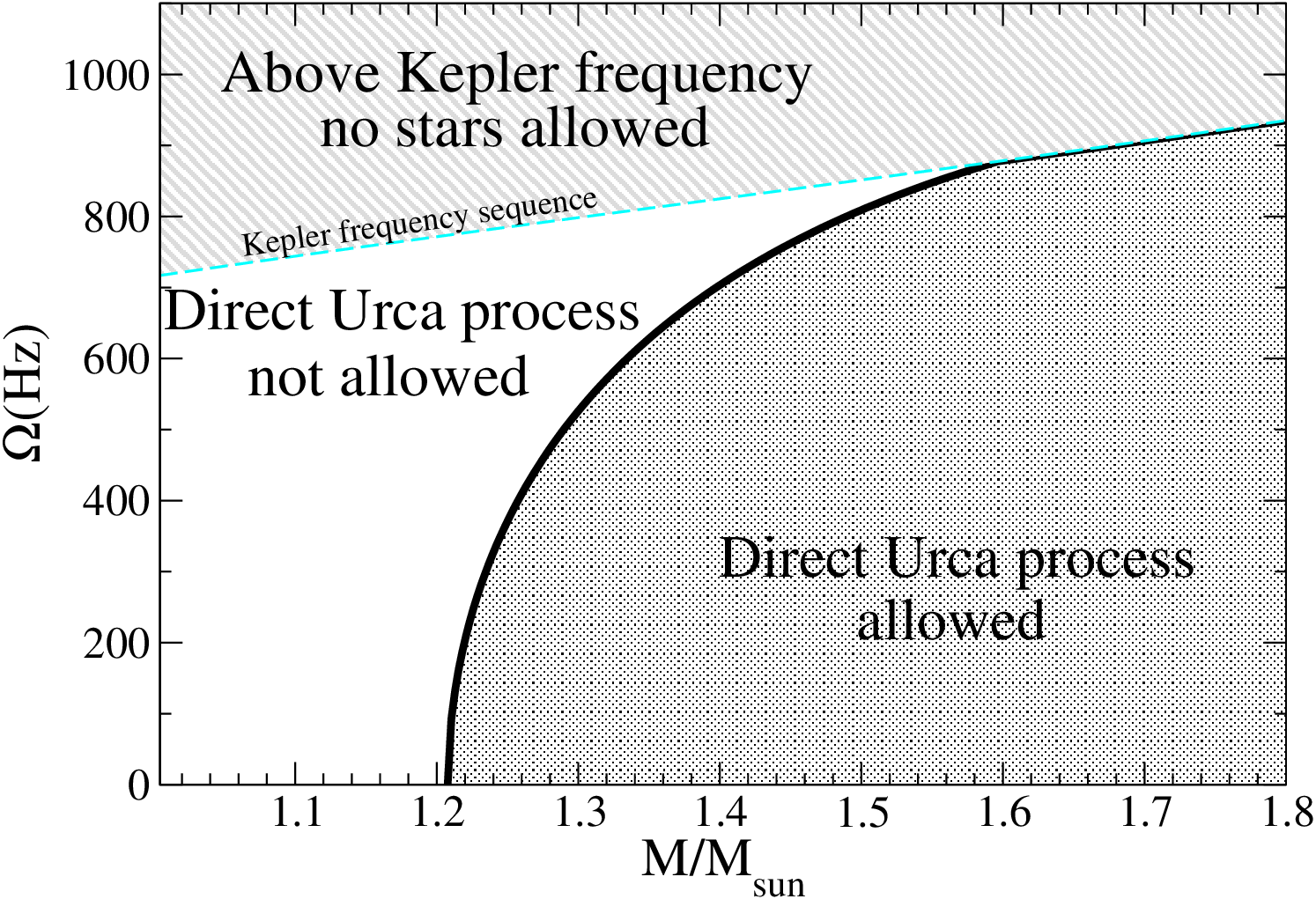}
\caption{\label{urca_region}(Color online) $\Omega(M)$ diagram. Any star
with mass and frequency that falls within the
dark gray shaded area allows for the direct Urca process to take place. The
light
grey shaded area, on the top of the diagram, represents a forbidden region,
where
no stars can be found since their frequency would be above their Kepler
frequency. Finally, the white area represents stars in which the direct Urca
process cannot take place. }
\end{figure}

Finally we have also generated a diagram $\Omega(M)$ that shows the domain of
frequencies and mass for which the direct Urca process is allowed. This result
is
shown in figure \ref{urca_region}, where any star with mass and frequency that
falls within the dark gray shaded area allows for the direct Urca process to
take place. The light grey shaded area, on the top of the diagram, represents a
forbidden region, where no stars can be found since their frequency would be
above their Kepler frequency. The white area represents stars in which
the direct Urca process cannot take place. Currently the spin of many X-ray
burst sources is known, and in some cases their masses can also be inferred (see
\cite{Ozel2006,Steiner2010,Physics2010} and references therein). In the case of
transient neutron stars, the core temperature can be inferred from the
quiescent emission state \cite{Heinke2007}. Furthermore, for some neutron
stars the spin frequency and the core temperature are bound (see for instance
\cite{Brown2002}). In the event that one might estimate the core temperature,
spin frequency and stellar mass, the diagram shown in Fig.\ \ref{urca_region}
might be used as a further test for this model.

\section{Conclusion}
A major advantage of our work, when compared to other studies of hybrid stars,
is that because we have only one equation of state for different degrees of freedom we
can study in detail the way in which chiral symmetry is restored and the way deconfinement
occurs in the stars. Such phenomena happens, for example, during the star spin down.

We have found that SU(3) non-linear sigma model is suitable  for the description of hybrid
stars. The predicted maximum masses and the respective radii lie in the observed
range. 
 For a static object, our
model predicts a star containing 2 km of hybrid matter (radius) surrounded by
hadronic matter. In the event that the object is rotating at its Kepler
frequency, the hybrid core becomes an oblate ellipsoid with equatorial and
polar
radii of 1.18 and 0.87 km, respectively. The reduction of the quark core for a
rotating object should not be surprising, if one considers the reduction in
density that follows rapid rotation.

We have also investigated the cooling of spherically symmetric (static) neutron
stars, whose composition is described by our model. We have found that the
threshold for the direct Urca process is reached for stars with mass greater
than 1.0 $M_\odot$. We have compared the cooling curves predicted by our model
with some prominent observed compact stars temperature. We have found that,
within our model, objects with masses up to 1.22 $M_\odot$ are in good
agreement with the observed data. Any star with mass above this value features
a thermal evolution that is too fast to be in agreement with the observed data.
This result might seem inconsistent with the
observed data, since most of these objects are expected to have a higher mass
than 1.22 $M_\odot$. It is important to notice that the results shown in
Fig. \ref{cooling} were obtained assuming a "froze-in" structure/composition.
As noted in reference \cite{Heinke2007}, many observed accreting neutron stars
are in agreement with the slow cooling scenario. The fact that they are
accreting implies that their structure is changing, as their frequency is
modified as a result of the accretion process.
As we showed in
Fig. \ref{Urca_rad}, the threshold radius for the direct Urca process strongly
depends on the frequency of the star. Therefore, the fast cooling for stars
with masses above 1.22 $M_\odot$ might be deceiving, since if we consider
rotation
we might obtain stars with $M_0 = 1.6 M_\odot$ without presenting the direct
Urca process (if the frequency is high enough). 
We could also see that a few observed objects present a very high temperature,
that cannot be explained by our model. Most likely there is some non-trivial
heating process taking place in these objects, which would explain why they are
so warm at their relatively old ages, this topic however is beyond the scope of this work.
It is important to mention that
the quarks found in the 2 km hybrid core, have little effect on the
cooling of these objects. Not only is the hybrid core very small, but the
quarks are present at a smaller ratio than the hadrons, which allows
the hadrons to dominate the cooling processes. On the other hand the suppression
of hyperons, due to the onset of the quark phase is important. Since the
hyperons do not appear in great quantities, some cooling channels (Hyperon Urca
processes for instance) are not open. These channels are not very
efficient cooling mechanisms \cite{Prakash1992}, nonetheless their absence
slow down the cooling.

\section{Acknowledgements}
We acknowledge the access to the computing facility of the Center of Scientific Computing
at the Goethe-University Frankfurt for our numerical calculations.

\end{document}